\documentclass[
 reprint,
 superscriptaddress,
 amsmath,
 amssymb,
 aps,
 prb
]{revtex4-2}

\usepackage[colorlinks=true,allcolors=blue]{hyperref}
\usepackage{graphicx}
\usepackage{dcolumn}
\usepackage{bm}
\usepackage{esint}
\usepackage{mathtools}

\usepackage[dvipsnames]{xcolor}

\begin{document}

\title{Reduction of fully screened magnetoplasmons in a laterally confined anisotropic two-dimensional electron system to an isotropic one}

\author{D.A. Rodionov}%
\email{rodionov.da@phystech.edu}
\affiliation{Kotelnikov Institute of Radioengineering and Electronics of Russian Academy of Sciences, Moscow, 125009 Russia}
\affiliation{Moscow Institute of Physics and Technology, Dolgoprudny, Moscow Region, 141701 Russia}

\author{I.V. Zagorodnev}%
\affiliation{Kotelnikov Institute of Radioengineering and Electronics of Russian Academy of Sciences, Moscow, 125009 Russia}

\date{\today}

\begin{abstract}
We investigate the properties of natural two-dimensional (2D) magnetoplasma modes in laterally confined electron systems, such as 2D materials, quantum wells, or inversion layers in semiconductors, with an elliptic Fermi surface. The conductivity of the system is considered in a dynamical anisotropic Drude model. The problem is solved in the fully screened limit, i.e., under the assumption that the distance between the two-dimensional electron system and the nearby metal gate is small compared to all other lengths in the system, including the wavelength of plasmons. Remarkably, in this limit plasma oscillations in an anisotropic 2D confined system are equivalent to plasma oscillations in an isotropic 2D electron system obtained by some stretching, even when the electromagnetic retardation is taken into account. Moreover, accounting for electromagnetic retardation leads only to a renormalization of the effective masses of carriers, somewhat like in relativity. As an example, we reduce the equations describing plasmons in a gated disk with an anisotropic two-dimensional electron gas to the equations describing oscillations in an isotropic ellipse. Without a magnetic field, we solve them analytically and find eigenfrequencies. To find a solution in a magnetic field, we expand the current of plasma oscillations in the complete set of Mathieu functions. Leaving the leading terms of the expansion, we approximately find and analyze magnetodispersion for the lowest modes.
\end{abstract}

\maketitle

\section{\label{sec:introduction} Introduction}

Plasmons, i.e., collective oscillations of charge carriers, in 2D electron systems (ESs) have been actively investigated over the last decades as they are of interest for applied physics, particularly in the generation and detection of sub- and terahertz radiation \cite{Shur2020, Elbanna2023}. An important advantage of 2D ESs is that the plasma frequency can be tuned by gate voltage or an external magnetic field.

Because the damping and quality factor of plasmons are determined primarily by electron collision decay rate, the observation of plasmons requires a system with sufficiently high mobility of carriers. Therefore, for a long time the main attention has been paid to plasmons in systems such as electrons on the surface of liquid helium, silicon field-effect transistor, and quantum wells based on, for example, Ga(Al)As \cite{Grimes1976,Allen1977}. These systems are isotropic in the plane, and, therefore, plasmons have been considered mainly in isotropic systems until recently.  Since then, however, the list of two-dimensional ES of sufficiently high quality has expanded considerably, and today, for example, much attention is being paid to plasmon or plasmon polariton in new 2D materials. A number of these systems, such as multilayer or monolayer black phosphorus \cite{Li2014,Rodin2014,Qiao2014,Low2014,Pogna2024}, the 1T$'$ phase of the transition metal dichalcogenides \cite{Lin2015, Chenet2015}, the trichalcogenides \cite{Dai2015, Saeed2017}, as well as the strained AlAs/AlGaAs quantum well \cite{Shayegan2006,Khisameeva2018,Khisameeva2020} are strongly anisotropic and the question of the properties of plasma oscillations in them has naturally arisen. It turns out that plasmons do not simply become anisotropic in these systems \cite{Low2014, Ahn2021, Sokolik2021,Miskovic2023}, but at some frequencies a new hyperbolic type of plasmons can also appear \cite{Nemilentsau2016, Ma2018}.

Plasma oscillations are much easier to analyze theoretically for laterally infinite systems. However, in practice they are excited in inhomogeneous samples, e.g., in laterally confined samples (strips, disks), where the theory of plasma oscillations is strongly complicated even in the case of the simplest description of 2D ESs in terms of the local Ohm's law and the Drude conductivity. Exact solutions for anisotropic confined systems were obtained only in a few cases, for example, for plasma oscillations in the half-plane in the quasi-stationary limit \cite{Margetis2020, Sokolik2021}. However, when the distance between the 2D ES and the large metal plate (gate) is much smaller than the characteristic lateral sizes of the 2D ES (the fully screened limit), the charge carriers in the 2D ES effectively interact only with their images under the gate and the analytical description of plasmons is drastically simplified \cite{Jin2016,Zagorodnev2023, Rodionov2023}. Note that this limit can be easily achieved \cite{Muravev2007, Bandurin2018, Alcaraz2018, Bylinkin2019}.

In our work, we study the general properties and eigenfrequencies of magnetoplasma modes in the fully screened 2D ESs with an anisotropic two-dimensional electron gas the Fermi surface of which is an ellipse. We describe plasmons in a classical way using Maxwell’s equations and the dynamical conductivity for the constitutive relation between the current density and the electric field in 2D ES. We are interested only in the frequency range in which the conductivity of the system is described by the (anisotropic) dynamical Drude model when the main contribution to the conductivity is determined by intraband transitions \cite{Nemilentsau2016}. This typically corresponds to the subterahertz and in some cases also terahertz frequency range. In Sec.~\ref{sec:equation} we describe key equations for the current density of plasma modes in the fully screened limit for an arbitrarily shaped 2D ES. In Sec.~\ref{sec:disk}, as an example, we solve the equations for an anisotropic disk and analyze the frequency, charge, and current distribution for the lowest plasma modes including a perpendicular magnetic field. Sec.~\ref{sec:conclusion}  presents concluding remarks.

\section{\label{sec:equation} Key equations for a freeform 2D ES}

\begin{figure}
    \includegraphics[width=\linewidth]{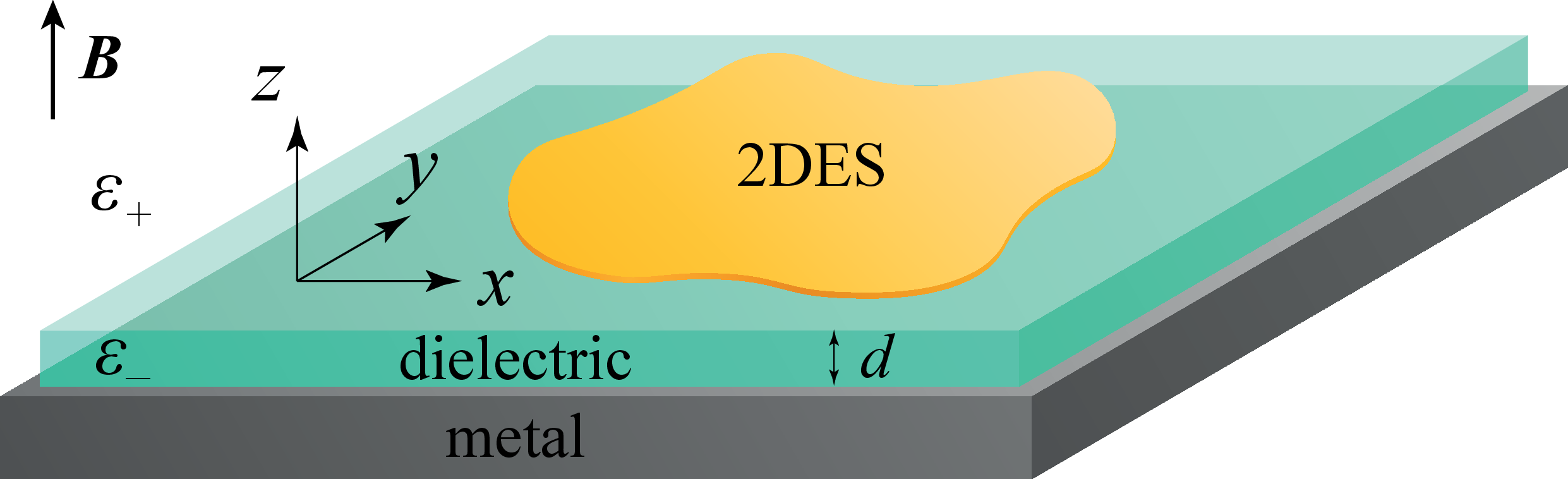}
    \caption{Schematic view of the system under consideration. The 2D ES is located in the plane $z=0$ between two dielectrics with permittivity $\varepsilon_{-}$ and $\varepsilon_{+}$. A perfectly conducting metal gate is under the two-dimensional electron gas. The system is placed in a magnetic field $\bm{B}$ being perpendicular to the 2D ES plane.}
    \label{fig:1}
\end{figure}

We study plasma oscillations in an anisotropic freeform 2D ES with an elliptic Fermi surface  (please, see Fig.~\ref{fig:1}). Above the 2D ES, the dielectric permittivity is $\varepsilon_{+}$. The 2D ES lies on a dielectric substrate with thickness $d$ and permittivity $\varepsilon_{-}$. Under the substrate is a flat perfectly conducting metallic gate. The entire system is placed in a perpendicular static homogeneous magnetic field $\bm{B}$. Let us introduce a Cartesian coordinate system such that the 2D ES and the gate are located in the $z=0$ and $z=-d$ planes, respectively. We denote by $S$ the region occupied by the 2D ES and by $\partial S$ its edge. Below we derive the main equation for the current density in the 2D ES and discuss the properties of the plasmons in the system.

We search for monochromatic plasma oscillations and, therefore, consider that their charge and current densities and induced electromagnetic fields are proportional to $e^{-i\omega t}$ where $\omega$ is the oscillation frequency. The expressions below are in the CGS. Using Maxwell's equations, it can be shown that the relationship between the tangential electric field $\bm{E}(\bm{r})$ and the current density $\bm{j}(\bm{r})$ in the 2D ES plane is given by the following expression:
\begin{multline}
    \bm{E}(\bm{r})=i\frac{4\pi}{\omega}\Big[(\nabla\otimes\nabla)\int\limits_{S}G_1(\bm{r}-\bm{r}')\bm{j}(\bm{r}')d\bm{r}'+\\
    \frac{\omega^2}{c^2}\int\limits_{S}G_2(\bm{r}-\bm{r}')\bm{j}(\bm{r}')d\bm{r}'\Big].
    \label{eq:connection_E_j}
\end{multline}
Here $\nabla$ is the 2D Nabla operator, $\otimes$ is the tensor product. The integral kernels $G_1$ and $G_2$ are given by the following expressions:
\begin{gather}
    G_1(\bm{r})=\int\limits_{\mathbb{R}^2}\frac{e^{i\bm{q}\cdot\bm{r}}}{\varkappa(q)Q(q)}\frac{d\bm{q}}{(2\pi)^2},\nonumber\\
    G_2(\bm{r})=\int\limits_{\mathbb{R}^2}\frac{e^{i\bm{q}\cdot\bm{r}}}{Q(q)}\frac{d\bm{q}}{(2\pi)^2},
    \label{eq:int_kernels}
\end{gather}
where
\begin{gather}
    \varkappa(q) = \frac{\varepsilon_{+}\beta_{-}\tanh(\beta_{-}d) + \varepsilon_{-}\beta_{+}}{\beta_{-}\tanh(\beta_{-}d) + \beta_{+}},\nonumber\\
    Q(q) = \beta_{+} + \beta_{-}\coth(\beta_{-}d),\nonumber\\
    \beta_{\pm} = \sqrt{q^2 - \varepsilon_{\pm}\frac{\omega^2}{c^2}},\quad\, \text{Re}\,\beta_{\pm} > 0.
    \label{eq:kappa_and_Q}
\end{gather}
The detailed derivation of this relation is presented in App. \ref{app:E_J_connection} and is similar to the technique used in Refs.~\cite{Mikhailov2005,Jin2016}. 

The next step to derive the equation for the current density is to use the local Ohm's law. Within the framework of this paper we will consider anisotropic 2D ESs with the following Drude-like conductivity:
\begin{gather}
    \sigma_{kk} = \frac{n_s e^2}{m_{k}}\frac{i\omega}{\omega^2-\omega_c^2},\quad k=x,y,\nonumber\\
    \sigma_{xy}=-\sigma_{yx}=\frac{n_s e^2}{\sqrt{m_x m_y}}\frac{\omega_c}{\omega^2-\omega_c^2},
    \label{eq:conductivity}
\end{gather}
where the axes $x$ and $y$ are chosen so that the effective mass tensor is diagonal, and the effective masses of electrons along them are $m_x$ and $m_y$, respectively. Here $-e$ and $n_s$ are the charge and the two-dimensional concentration of electrons, $\omega_c=e B_z/\sqrt{m_x m_y}c$ is the cyclotron frequency. We assume that the system is ``clean'', i.e., the plasma frequency and cyclotron frequency are much greater than the inverse scattering time of the electrons in the system. Also, we require the normal component of the current density is equal to zero at the edge of the system.

The obtained integro-differential equation is hardly solvable analytically. However, we consider the case of strong screening when the distance $d$ between the metal and 2D ES is small compared to the characteristic lengths in the system \cite{Fetter1986,Volkov1988}. In this limit, we decompose the integral kernels up to first nonvanishing order in the distance $d$ and have $\varkappa(q) \approx \varepsilon_{-}$, $1/Q(q) \approx d$. In this case, the kernels (\ref{eq:int_kernels}) reduce to the Dirac delta functions: $G_1(\bm{r}) = d\,\delta(\bm{r})/\varepsilon_{-}$ and $G_2(\bm{r}) = d\,\delta(\bm{r})$. Thus, in the fully screened limit, the equation for the current density has the following form:
\begin{equation}
    \bm{j}(\bm{r}) = i\frac{4\pi d\sigma}{\varepsilon_{-}\omega}\left[\nabla\otimes\nabla + \varepsilon_{-}\frac{\omega^2}{c^2}I\right]\bm{j}(\bm{r}),
    \label{eq:main}
\end{equation}
where $I$ is an identity matrix. Note that, as one would expect, in this limit the charges and currents interact with their images in the metal locally and, therefore, only the dielectric permittivity of the substrate remains. For brevity, we omit the ``$-$'' index of the dielectric permittivity below. Now we turn to analysis of the Eq. (\ref{eq:main}).

First we show that the problem of the study of plasma oscillations taking into account the electromagnetic retardation can be reduced to the study of plasma oscillations in the quasi-stationary limit in a 2D ES of the same shape with another effective conductivity tensor. To demonstrate this, we note that the electromagnetic retardation takes place only in the term proportional to $\omega^2/c^2$. Thus, Eq.~(\ref{eq:main}) can be rewritten as
\begin{gather}
    \bm{j}(\bm{r}) = i\frac{4\pi d\sigma^{*}}{\varepsilon\omega} \left(\nabla\otimes\nabla\right) \bm{j}(\bm{r}),\nonumber\\
    \sigma^{*}=\left(\sigma^{-1}-i\frac{4\pi d\omega}{c^2}I\right)^{-1}
    \label{eq:j_short}
\end{gather}
with the same boundary condition. In other words, if in the quasi-stationary limit we want to take into account the electromagnetic retardation, it is necessary to ``shorten'' the diagonal elements of the resistivity tensor $\sigma^{-1}$ by the value $i4\pi d\omega/c^2$. The obtained property is fulfilled for any conductivity tensor. Note that the transition to the effective tensor $\sigma^{*}$ is equivalent to increasing the effective masses $m_{i}$ ($i=x,y$) by the value $\delta m=4\pi d n_s e^2/c^2$ in the original conductivity tensor $\sigma$, including the definition of $\omega_c$. We denote the new masses by $m_x^{*}$ and $m_y^{*}$, respectively, and the new cyclotron frequency as $\omega_c^*$.

Now we show that the plasma oscillations in an anisotropic system can be reduced to the oscillations of an isotropic system by some stretching and shrinking. Let us transform the coordinates as follows
\begin{equation}
    \tilde{\bm{r}} = T \bm{r},\quad
    T = 
    \begin{pmatrix}
        \sqrt[4]{\frac{m_x^*}{m_y^*}} & 0\\
        0 & \sqrt[4]{\frac{m_y^*}{m_x^*}}
    \end{pmatrix}.
    \label{eq:T}
\end{equation}
In this case, the Nabla operator is transformed as $\nabla=T\tilde{\nabla}$. Using the diagonal form of the matrix $T$ and the properties of the tensor product, the equation (\ref{eq:j_short}) can be written in the following form:
\begin{equation}
    \tilde{\bm{j}}(\tilde{\bm{r}}) = i\frac{4\pi d\tilde{\sigma}}{\varepsilon\omega} \left(\tilde{\nabla}\otimes\tilde{\nabla}\right)\tilde{\bm{j}}(\tilde{\bm{r}}),
    \label{eq:j_isotropic}
\end{equation}
where $\tilde{\sigma}=T\sigma^* T$ and $\tilde{\bm{j}}(\tilde{\bm{r}})=T\bm{j}(T^{-1}\tilde{\bm{r}})$. The trick is that the transformed tensor $\tilde{\sigma}$ is isotropic with isotropic effective mass $m^*=\sqrt{m_x^* m_y^*}$. 

As a last step, let us show explicitly that the component of the transformed current normal to the boundary is zero, i.e., the boundary condition does not change. We introduce a parameterization of the boundary $\partial S=\{\bm{r}(\xi)\,|\,\xi\in [0,1)\}$, where $\bm{r}(0)=\bm{r}(1)$ and the increase of $\xi$ corresponds to counterclockwise contour traversal. The external normal $\bm{n}$ to the edge is given by the following expression:
\begin{equation}
    \bm{n} = R \frac{\bm{r}'(\xi)}{|\bm{r}'(\xi)|}, \quad 
    R = 
    \begin{pmatrix}
        0 & -1\\
        1 & 0
    \end{pmatrix},
\end{equation}
where $\bm{r}'(\xi)$ is the derivative. After coordinate transformation, the normal can be expressed through the normal $\tilde{\bm{n}}$ to the contour $\tilde{\bm{r}}(\xi)$:
\begin{equation}
    \bm{n}=\frac{|\bm{r}
'(\xi)|}{|T^{-1}\bm{r}'(\xi)|}R T^{-1} R^{-1}\tilde{\bm{n}},
\end{equation}
where the explicit substitution of the matrices $R$ and $T$ gives the equality $R T^{-1} R^{-1}=T$. Taking advantage of the fact that $T$ is a diagonal matrix, the properties of the scalar product and the relation between normals obtained above, the boundary condition of the original problem reduces to the boundary condition $\tilde{\bm{n}}\cdot\tilde{\bm{j}}(\tilde{\bm{r}})\big|_{\tilde{\bm{r}}\in\partial\tilde{S}}=0$, where $\partial \tilde{S}=\{\tilde{\bm{r}}(\xi)\,|\,\xi\in [0,1)\}$. Thus, the boundary condition retains its form.

Thus, we have reduced the dynamics of electrons in an anisotropic system with electromagnetic retardation to an isotropic system without retardation. It is convenient to introduce a scalar potential $\varphi$, substituting the current density in the form $\tilde{\bm{j}}(\tilde{\bm{r}}) = -\tilde{\sigma}\tilde{\nabla}\varphi(\tilde{\bm{r}})$. This potential is indeed electric potential only in the quasistatic limit. Then the equation Eq.~(\ref{eq:j_isotropic}) and its boundary condition can be reduced to the following form:
\begin{gather}
    \tilde{\Delta}\varphi(\tilde{\bm{r}})=-\frac{\omega^2-{\omega_c^*}^2}{{v^*}^2}\varphi(\tilde{\bm{r}}),\nonumber\\
    \left[\frac{\partial}{\partial\tilde{n}} + i\frac{\omega_c^*}{\omega}\frac{\partial}{\partial\tilde{\tau}}\right]\varphi(\tilde{\bm{r}})\big|_{\tilde{\bm{r}}\in\partial \tilde{S}}=0,
    \label{eq:potential}
\end{gather}
where ${v^*}^2 = 4\pi d n_s e^2/\varepsilon m^*$, $\tilde{\bm{\tau}} = R\tilde{\bm{n}}$ is the tangent to the edge $\partial\tilde{S}$, and derivatives along the directions $\partial/\partial\tilde{n} = \tilde{\bm{n}}\cdot\tilde{\nabla}$ and $\partial/\partial\tilde{\tau} = \tilde{\bm{\tau}}\cdot\tilde{\nabla}$. The resulting system of equations is the key equations under consideration.

Now let us discuss general properties of solutions of these key equations. In the Hilbert space of functions satisfying the boundary condition Eq.~(\ref{eq:potential}), with scalar product $\langle f|g \rangle = \int_{\tilde{S}}\overline{f}(\bm{r})\cdot g(\bm{r})d\bm{r}$, the Laplace operator is Hermitian \cite{Gabov1976}. As a consequence, the eigenvalues are real and the corresponding eigenfunctions can be indexed by two numbers, which we combine in the vector $\bm{n}$. In addition, the eigenvalues depend only on the parameter $\omega_c/\omega$ included in the boundary condition and the geometry of the boundary $\partial\tilde{S}$. Thus, if $-k^2_{\bm{n}}(\omega_c^*/\omega,\partial\tilde{S})$ denotes the eigenvalue corresponding to the eigenfunction with numbers $\bm{n}$, then the dispersion equation for the frequency of the plasma mode is as follows
\begin{equation}
    \omega^2 - {\omega_c^*}^2 = {v^*}^2\,k^2_{\bm{n}}\left(\frac{\omega_c^*}{\omega},\partial\tilde{S}\right).
\end{equation}
It can be seen that there exists a solution with everywhere constant potential and zero eigenvalue. However, this mode is of no interest since its current and charge density are zero. In the absence of a magnetic field, we have the Neumann boundary condition for the sought eigenfunctions. In this case, the eigenvalues $-k_{\bm{n}}^2(0,\partial\tilde{S})$ are not positive. The frequency of plasma modes is $v^*\,k_{\bm{n}}(0,\partial\tilde{S})$, where the value $k_n(0,\partial\tilde{S})$ has the inverse length dimension and actually defines the rule of quantization of the wave vector in a laterally confined 2D ES. Note that in the presence of a magnetic field, if the value $k^2_{\bm{n}}(\omega_c^*/\omega,\partial\tilde{S})$ takes a negative value in the frequency region $\omega<\omega_c^*$, then there should exist solutions corresponding to edge magnetoplasmons \cite{Volkov1985,Mast1985,Jin2016}.

As a final remark, note that the reduction of an anisotropic system to an isotropic one is a feature of the fully screened limit. Indeed, the transformation $T$ stretches the original 2D ES along one direction and compresses along the other direction by the same value, and hence similarly transforms the interaction between charges, making it anisotropic. However, in the fully screened limit, the interaction potential is local and is expressed by the Dirac delta functions. Since $\delta(\bm{r})=\delta(x)\delta(y)$ and $\delta(\alpha x)=\delta(x)/|\alpha|$ for a real number $\alpha$, the transformation $T$ in the kernels does not change them only in the fully screened limit.

\section{\label{sec:disk} Plasmons in an anisotropic disk}

To illustrate the established properties, consider a 2D ES in the form of a disk of radius $R$. In an equivalent isotropic system, the boundary of 2D ES is transformed into an ellipse with semi-major $a = R\sqrt[4]{m_x^*/m_y^*}$ and semi-minor $b = R\sqrt[4]{m_y^*/m_x^*}$ axes and focus $f=\sqrt{a^2-b^2)}$. Without loss of generality, we assume $m_x > m_y$ and, correspondingly, $m_x^*>m_y^*$. Let the origin of the coordinate system $(x,y)$ be located at the center of the disk, and hence of the ellipse. To solve Eqs.~(\ref{eq:potential}) inside the ellipse, we introduce the elliptic coordinates:
\begin{equation}\label{eq:el_coords}
    \begin{cases}
        \tilde{x}=f\cosh\mu\cos\nu,\\
        \tilde{y}=f\sinh\mu\sin\nu,
    \end{cases}
\end{equation}
where $\nu\in[0,2\pi)$ and $\mu\ge 0$, and Eq.~(\ref{eq:potential}) can be written in the following form:
\begin{gather}
    \left(\frac{\partial^2}{\partial\mu^2} + \frac{\partial^2}{\partial\nu^2}\right)\varphi(\mu,\nu)=2\xi(\omega_c^*)(\cos 2\nu-\cosh 2\mu)\varphi(\mu,\nu),\nonumber\\
    \left[\frac{\partial}{\partial\mu} + i\frac{\omega_c^*}{\omega}\frac{\partial}{\partial\nu}\right]\varphi(\mu,\nu)\Big|_{\mu=\mu_0}=0.
    \label{eq:potential_ell}
\end{gather}
Here $\xi(\omega_c^*) = (\omega^2-{\omega_c^*}^2)f^2/(2v^*)^2$ and $\cosh\mu_0 = a/f$.

\subsection{Absence of a magnetic field}

Using the separation of variables, we solve Eq.~(\ref{eq:potential_ell}) in the absence of a magnetic field. Detailed mathematical calculations are given in App. \ref{app:solve_B=0}.

\begin{figure}
    \includegraphics[width=\linewidth]{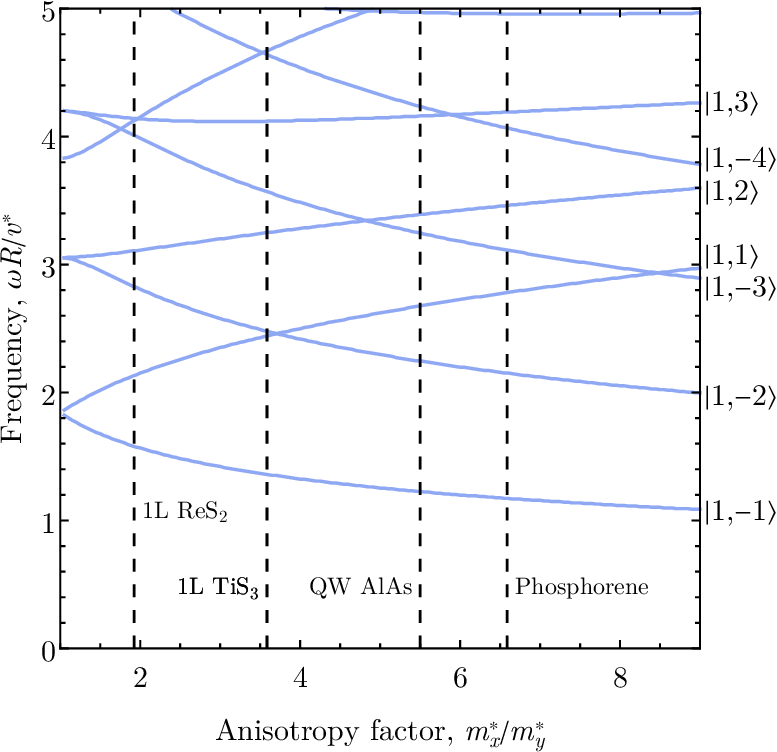}
    \caption{Dependence of the eigen frequency of plasma oscillations in the disk on the ratio of effective masses in the absence of a magnetic field. The dashed lines correspond to different anisotropic systems (in the quasi-stationary limit $m_x^*=m_x,m_y^*=m_y$): the AlAs/AlGaAs quantum well with $m_x=1.1$, $m_y=0.2$ \cite{Shayegan2006}, phosphorene with $m_x=1.12$, $m_y=0.17$ \cite{Qiao2014}, a monolayer (1L) of TiS$_3$ with $m_x=1. 47$, $m_y=0.41$ \cite{Dai2015}, and a monolayer of ReS$_2$ with $m_x=0.25$, $m_y=0.13$ \cite{Lin2015}. Accounting for electromagnetic retardation will decrease the ratio $m_x^*/m_y^*$, therefore the vertical dashed lines will shift to the left.}
    \label{fig:2}
\end{figure}

\begin{figure*}
    \includegraphics[width=\linewidth]{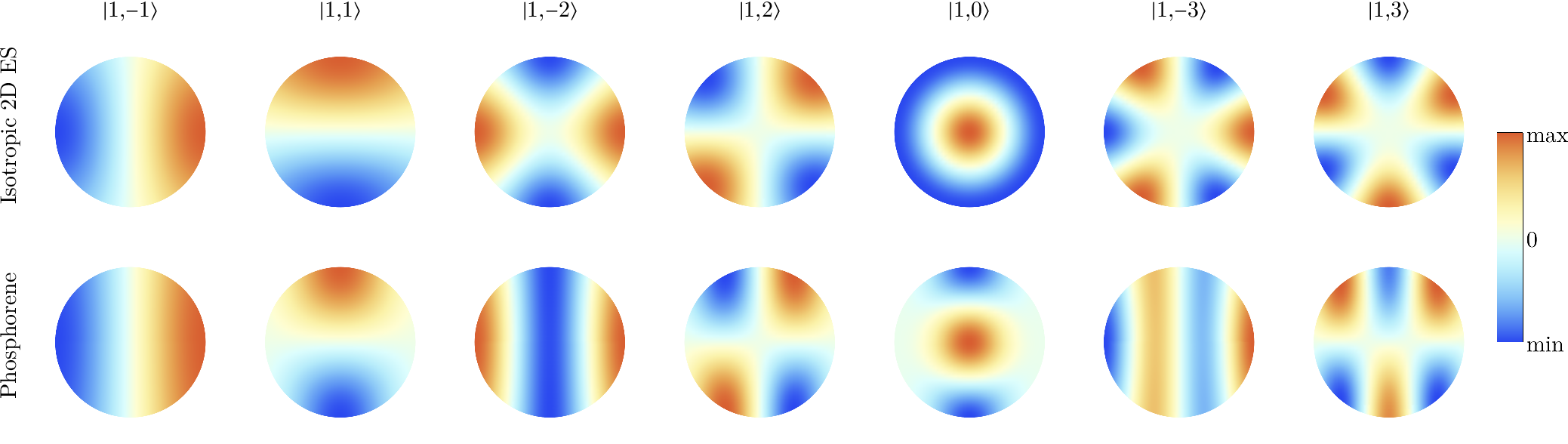}
    \caption{Charge density of seven lowest plasma modes in an isotropic (top) and anisotropic (bottom) 2D ES. The effective masses for the anisotropic system were taken for phosphorene: $m_x^*=m_x=1.12$, $m_y^*=m_y=0.17$.}
    \label{fig:3}
\end{figure*}

The natural plasma modes can be numbered by two integers, which we call elliptic radial $n=1,2,...$ and azimuthal $m=0,\pm 1,\pm 2,...$ numbers. The scalar potential of the plasma mode is given by the following expression:
\begin{gather}
    \varphi_{n,m}(\mu,\nu)=C_{n,m}M_{m}\left(\mu,\frac{\omega_{n,m}^2f^2}{4{v^*}^2}\right)N_{m}\left(\nu,\frac{\omega_{n,m}^2f^2}{4{v^*}^2}\right),\nonumber
    \\
    M_{m}(\mu,\xi)=
    \begin{cases}
        \text{Se}_m(\mu,\xi), & m > 0,\\
        \text{Ce}_{|m|}(\mu,\xi), & m \le 0,
    \end{cases}\nonumber
    \\
    N_{m}(\nu,\xi)=
    \begin{cases}
        \text{se}_m(\nu,\xi), & m > 0,\\
        \text{ce}_{|m|}(\nu,\xi), & m \le 0,
    \end{cases}
    \label{eq:plasmon_potential}
\end{gather}
where $\text{ce}_m(\nu,\xi)$ and $\text{se}_m(\nu,\xi)$ are even and odd Mathieu functions, respectively, and $\text{Ce}_m(\mu, \xi) = \text{ce}_m(i\mu,\xi)$ and $\text{Se}_m(\mu,\xi) = -i\,\text{se}_m(i\mu,\xi)$ are modified Mathieu functions of the first kind. The normalization coefficients $C_{n,m}$ can be chosen from the considerations that the basis of the functions $\varphi_{n,m}$ is orthonormalized. The natural frequency $\omega_{n,m}$ is given by the equality
\begin{equation}\label{eq:frequency_B=0}
    \omega_{n,m}=v^*\frac{\alpha_{n,m}}{R},
\end{equation}
where the coefficient $\alpha_{n,m}$, specifying the quantization rule of the wave vector, satisfies the equation
\begin{equation}\label{eq:alpha}
    \alpha_{n,m}\sqrt[4]{\frac{m_x^*}{m_y^*}} = \eta_{m}^{(n)}\left(\frac{\alpha_{n,m}^2}{4}\left[\sqrt{\frac{m_x^*}{m_y^*}}-\sqrt{\frac{m_y^*}{m_x^*}}\right]\right)
\end{equation}
and depends only on the mass ratio. Here $\eta_{m}^{(n)}(\xi)$ is the $n$-th nontrivial zero of the derivative of the function $M_m(\text{arcosh}(\eta/2\sqrt{\xi}),\xi)$ on the parameter $\eta$. In the isotropic case, $\eta_{m}^{(n)}(0)$ is equal to the $n$-th nontrivial zero of the derivative of the Bessel function of the first kind of order $|m|$.

Below we will use the Dirac formalism and denote the plasma mode with numbers $n$ and $m$ as $|n,m\rangle$.

In the isotropic case, the plasma modes $|n,m\rangle$ and $|n,-m\rangle$ are degenerate at $m\neq 0$, and any linear combination of them can be realized. In the literature discussing plasmon oscillations in an isotropic disk, one often uses an azimuthal number $l$ ($l=0,\pm 1,\pm 2,...$) such that the charge and current densities of plasmons $\propto e^{il\nu}$ \cite{Fetter1986}. In the case, the mode described through the azimuthal number $l\neq 0$ is constructed as $(|n,|m|\rangle \pm i|n,-|m|\rangle)/\sqrt{2}$, where ``$\pm$'' is defined, for example, by the sign $l$.

In Fig.~\ref{fig:2} we present the dependence of the frequency of the lowest plasma modes on the mass ratio $m_x^*$ and $m_y^*$ obtained from Eq.~(\ref{eq:frequency_B=0}). In an isotropic system all plasma modes, except $m=0$, are degenerate. Anisotropy results in their splitting. To understand qualitatively how this splitting occurs, consider the $|1,\pm 1\rangle$ plasma modes. In the plasma mode $m=1$, the charge oscillation occurs mainly along the minor axis, while in the mode $m=-1$ it occurs along the major one. Thus, the effective size-quantized wave vector, which is inverse proportional to the distance between oscillating charges, is greater for the $|1,1\rangle$ mode than for the $|1,-1\rangle$ mode. Since in the fully screened limit the frequency is directly proportional to the effective wave vector, Eq.~(\ref{eq:frequency_B=0}), there is the splitting of these modes.

In Fig.~\ref{fig:2} we mark popular 2D ESs such as a monolayer of TiS$_3$, a quantum well AlAs/AlGaAs, black phosphorus (phosphoren) monolayer, and a monolayer of TMD with dashed lines. These lines correspond to the quasi-stationary case when $m_i^*=m_i$. As the influence of electromagnetic retardation increases, the $\delta m$ term included in the definition of the $m_i^*$ masses increases and the ratio $m_x^*/m_y^*$ monotonically decreases from the value of $m_x/m_y$ in the quasi-stationary limit. Thus, as the influence of electromagnetic retardation increases, the dashed lines shift to the left and can lead to a change in the order of mode frequencies.

In Fig.~\ref{fig:3} we depict the charge density of the lowest plasma modes in an isotropic disk and in a disk of phosphorene. Because of the strong anisotropy in phosphorene, the equivalent ellipse with isotropic conductivity is strongly stretched along the major semi-axis and compressed along the minor semi-axis so that it can be thought of as an elongated strip. That is why the oscillation pattern of plasma modes with negative numbers $m$ has a stripe structure in the charge density peculiar to rectangular geometry.

\subsection{Magnetoplasmons}

In the presence of a magnetic field, we expand the scalar potential in a complete set of Mathieu functions as follows:
\begin{equation}
    \varphi(\mu,\nu) = \sum_{m'}C_{m'}M_{m'}(\mu,\xi(\omega_c^*))N_{m'}(\nu,\xi(\omega_c^*)),
    \label{eq:potential_Bneq0}
\end{equation}
where $m = 0,\pm 1,\pm 2,...$, the summation is over all integers and the functions $M_m$ and $N_m$ are defined in Eq.~(\ref{eq:plasmon_potential}). Each term of the series satisfies Eq.~(\ref{eq:potential_ell}).

Then we substitute the expansion into the boundary condition Eq.~(\ref{eq:potential_ell}), consecutively multiply it by the basis functions and integrate it over the variable $\nu$ to obtain the following system of linear equations for the unknown coefficients $C_m$:
\begin{equation}
        \sum_{m'}\left(A_{m,m'} + i\frac{\omega_c^*}{\omega}B_{m,m'}\right)C_{m'}=0,
        \label{eq:system_Bne0}
\end{equation}
\begin{gather}
    A_{m,m'} = \pi\frac{d}{d\mu}M_{m}(\mu,\xi(\omega_c^*))\Big|_{\mu=\mu_0}\delta_{m,m'},\\
    B_{m,m'} = \nonumber \\
    M_{m'}(\mu_0,\xi(\omega_c^*))
    \int\limits_{0}^{2\pi}N_{m}(\nu,\xi(\omega_c^*))\frac{d}{d\nu}N_{m'}(\nu,\xi(\omega_c^*))d\nu.
    \label{eq:matrix_Bne0}
\end{gather}
The solvability condition $\text{det}\,\left(A + i\frac{\omega^{*}_{c}}{\omega} B\right) = 0$ determines the frequency of the natural plasma modes. Due to the properties of Mathieu functions, the resulting system of linear equations decomposes into two subsystems. The first one describes the relationship of coefficients with even indices, and the second one with odd indices. More details are presented in App.~\ref{app:solve_Bneq0}.

\begin{figure}
    \includegraphics[width=0.9\linewidth]{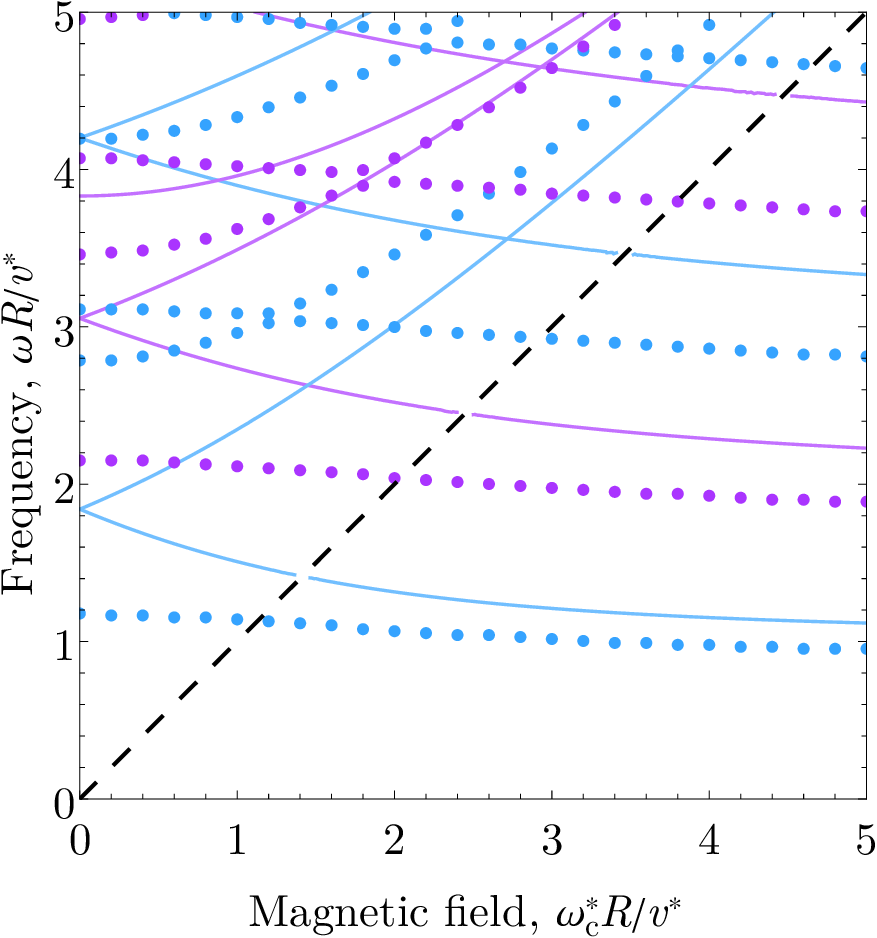}
    \caption{Dependence of the frequency of plasma oscillations in the disk on the magnetic field. The dots indicate the phosphorene magnetodispersion ($m_x=1.12$, $m_y=0.17$) \cite{Qiao2014}. Solid lines correspond to magnetodispersion in isotropic systems. The colors of the dots and solid lines indicate the different subsystems from which they are derived. The dashed line shows the frequencies $\omega=\omega_c^*$.}
    \label{fig:4}
\end{figure}

In Fig.~\ref{fig:4} we present the magnetodispersion, i.e., the dependence of the frequency on the magnetic field, for phosphorene. The solutions corresponding to different subsystems are indicated by different colors. As in the isotropic disk, there are ``bulk'' plasma modes whose magnetodispersion approaches the frequency $\omega_c^*$ and edge magnetoplasmons whose frequency tends to a constant value at the strong magnetic field \cite{Zagorodnev2023}. Note, the effective cyclotron frequency $\omega_c^*$ is smaller than the cyclotron frequency $\omega_c$, what is shown for an infinite system in \cite{Zabolotnykh2021}. As can be seen from the figure, the frequency of the edge magnetoplasmon depends on the magnetic field  weakly. In the case of strong anisotropy, the equivalent system is an elongated isotropic ellipse, which in some approximation can be considered as a long rectangle. And in such geometry the frequency of the edge magnetoplasmon depends weakly on the magnetic field \cite{Rodionov2024}. In addition, we can observe anticrossing (avoided crossing) of the magnetodispersion of some modes. It is related to the breaking of circular symmetry \cite{Zarezin2023,Rodionov2024}, which in our case is caused by anisotropy in the system.

\section{\label{sec:conclusion} Conclusion and discussion}

In our work, we have considered the general properties of magnetoplasma oscillations in fully screened anisotropic 2D ESs in which the Fermi surface is an ellipse and the conductivity can be considered in a dynamic anisotropic Drude model. We have shown that the electromagnetic retardation in the electron system can be taken into account in the quasi-stationary limit by ``increasing'' the effective electron masses by the value $\delta m = 4\pi d n_s e^2/c^2$ in the expression for the conduction tensor. At the same time, the study of the properties of plasma oscillations in such a system can be reduced to their study in a stretched system with isotropic Drude conductivity, but the effective mass is replaced by the geometric mean of the increased masses.

The fully screened limit implies that the distance between the 2D ES and the gate is small compared to all characteristic lengths in the system. Thus, in the absence of a magnetic field, the wavelength of the plasma oscillations, which is given by the expression $2\pi/k_n(0,\tilde{S})$, has to be much greater than the distance $d$. In the presence of a magnetic field, there is another characteristic dimension --- the localization length of the edge magnetoplasmon. As the magnetic field increases, the length decreases \cite{Volkov1988,Rodionov2023}. Therefore, the magnetic field should not be too large.

Let us conclude by assessing the impact of the electromagnetic retardation. Using convenient unit dimensions, the additive mass is
\begin{equation*}
    \delta m/m_0 = 3.5\times 10^{-4} n_s[10^{12}\,\text{cm}^{-2}] d[\mu\text{m}],
\end{equation*}
where $m_0$ is a free electron mass. Thus, for the additive to be comparable to the effective mass of the charge carrier, $m_0$, a sufficiently high carrier concentration or a sufficiently distant gate is required. However, to remain within the fully screened approximation, the latter requires large lateral lengths of the sample, millimeters and more for typical 2D ESs, which is realistic  for quantum wells and 2D materials \cite{Gusikhin2018,Chen2023}. Thus, in modern structures the electromagnetic retardation may sometimes be important.


\section{Acknowledgments}

This work was supported by the Foundation for the Advancement of Theoretical Physics and Mathematics ``BASIS'' (Project No. 21-1-5-133-1). We are grateful to V. M. Muravev, A. R. Khisameeva, and A.A. Zabolotnykh for valuable discussions.

\appendix

\section{\label{app:E_J_connection} Relation between electric field and curent density}

Here we establish the relation between the electric field in the 2D ES plane and the current density generating it. We consider the charge and current densities, and the electromagnetic fields induced by them are proportional to $e^{-i\omega t}$, where $\omega$ is the oscillation frequency. In this case, Maxwell's equations take the form:
\begin{gather}
    \text{rot} \bm{E}_{3D}(x,y,z) = i\frac{\omega}{c}\bm{H}_{3D}(x,y,z),\label{eq:Maxwell_rotE}\\
    \text{rot} \bm{H}_{3D}(x,y,z) = \frac{4\pi}{c}\bm{J}(x,y)\delta(z)-i\frac{\omega}{c}\varepsilon(z)\bm{E}_{3D}(x,y,z),\label{eq:Maxwell_rotH}
\end{gather}
where $\bm{E}_{3D}$ and $\bm{H}_{3D}$ are the electric and magnetic fields, $\rho$ and $\bm{J}$ are the charge and current densities in the two-dimensional electron system, $\bm{J}=(\bm{j},0)^T$, $\delta(z)$ is the Dirac delta function. By $\varepsilon(z)$ we denote the dielectric constant of the medium, i.e., $\varepsilon(z<0)=\varepsilon_{-}$ and $\varepsilon(z>0)=\varepsilon_{+}$. Eventually, from the total electric field $\bm{E}_{3D}=(\bm{E},E_z)^T$ we are going to be interested only in its tangential component $\bm{E}$ in the 2D ES plane $z = 0$. Using Eqs.~(\ref{eq:Maxwell_rotE}) and (\ref{eq:Maxwell_rotH}), we obtain the following equation:
\begin{equation}\label{eq:DE=j}
    \left[\text{rot\,rot}-\varepsilon(z)\frac{\omega^2}{c^2}\right]\bm{E}_{3D}(x,y,z)=i\frac{4\pi\omega}{c^2}\bm{J}(x,y)\delta(z).
\end{equation}
To express the electric field through the current density, it is necessary to find the inverse to the differential operator acting on the electric field strength. In this case, the inverse operator must take into account that the tangential component of the field is continuous and is zero on the metal plane $z=-d$. Applying the Fourier transform on coordinates $x$ and $y$ to the equation Eq.~(\ref{eq:DE=j}), we obtain:
\begin{multline}
    \begin{pmatrix}
        -\bm{q}\otimes\bm{q} + \left(q^2-\frac{\partial^2}{\partial z^2}-\varepsilon(z)\frac{\omega^2}{c^2}\right)I & i\bm{q}\frac{\partial}{\partial z}\\
        i\bm{q}^T\frac{\partial}{\partial z} & q^2-\varepsilon(z)\frac{\omega^2}{c^2}
    \end{pmatrix}\cdot\\
    \begin{pmatrix}
        \bm{E}(\bm{q},z)\\
        E_z(\bm{q},z)
    \end{pmatrix}
    =i\frac{4\pi\omega}{c^2}\bm{j}(\bm{q})\delta(z).
\end{multline}
where $\bm{q}$ is the Fourier transform parameter (the wave vector) and $I$ is a unit matrix $2\times 2$. From the third line of the equation we express the component $E_z$ and substitute it into the first two lines:
\begin{multline}\label{eq:DE=j_Fourier}
    \Bigg[-\frac{\partial^2}{\partial z^2} + \left(\bm{q}\otimes\bm{q}\right)\frac{\partial}{\partial z}\frac{1}{q^2-\varepsilon(z)\frac{\omega^2}{c^2}}\frac{\partial}{\partial z}+\\
    \left(q^2 I-\bm{q}\otimes\bm{q}-\varepsilon(z)\frac{\omega^2}{c^2}I\right)\Bigg]\bm{E}(\bm{q},z)=\\
    i\frac{4\pi\omega}{c^2}\bm{j}(\bm{q})\delta(z).
\end{multline}

As a result, we obtain an ordinary differential equation. To solve it, let us first consider the electric field separately in space $z>0$ and $z<0$. After some algebra, the equation can be rewritten in the following form:
\begin{equation}
    \left(-I+\frac{\bm{q}\otimes\bm{q}}{\beta_{i}^2}\right) \left(\frac{\partial^2}{\partial z^2} - \beta_{i}^2\right)\bm{E}_i(\bm{q},z) = 0,
\end{equation}
where the index $i$ is ``$+$'' or ``$-$'' and indicates the region of space $z>0$ or $z<0$, respectively. Here we introduce the notation $\beta_{i}^2=q^2-\varepsilon_{i}\frac{\omega^2}{c^2}$. The solution to the resulting equation is a linear combination of the functions $e^{\beta_i z}$ and $e^{-\beta_i z}$. Looking for a continuous function such that it is decreasing at $z\rightarrow\infty$ and zero in the plane $z=-d$, we arrive at the following solution:
\begin{equation}\label{eq:E_scheme}
    \bm{E}(\bm{q},z)=\bm{E}_0(\bm{q})
    \begin{cases}
         e^{-\beta_{+}z}\sinh\beta_{-}d, & z\ge 0,\\
        \sinh\beta_{-}(z+d), & z<0,
    \end{cases}
\end{equation}
where the branch of the root in the definition of $\beta_{+}$ is chosen such that $\text{Re}\,\beta_{+}>0$. The unknown coefficient $\bm{E}_0$ is determined from the condition dictated by the presence of the Dirac delta function in Eq.~(\ref{eq:DE=j_Fourier}):
\begin{multline}
    \left(-I+\frac{\bm{q}\otimes\bm{q}}{\beta_{+}^2}\right)\frac{\partial\bm{E}(\bm{q},z)}{\partial z}\Big|_{z=+0} - \\
    \left(-I+\frac{\bm{q}\otimes\bm{q}}{\beta_{-}^2}\right)\frac{\partial\bm{E}(\bm{q},z)}{\partial z}\Big|_{z=-0}=i\frac{4\pi\omega}{c^2}\bm{j}(\bm{q}).
\end{multline}
Expressing $\bm{E}_0$, substituting it into Eq.~(\ref{eq:E_scheme}) and considering the plane $z=0$, we obtain the relationship between the Fourier images of the electric field and the current density in the 2D ES plane:
\begin{gather}
    \bm{E}(\bm{q}) = i\frac{4\pi}{\varkappa(q)\omega}\left(-\bm{q}\otimes\bm{q} + \varkappa(q)\frac{\omega^2}{c^2}\right)\frac{1}{Q(q)}\bm{j}(\bm{q})
\end{gather}
where $\varkappa(q)$ and $Q(q)$ are defined in (\ref{eq:kappa_and_Q}). Applying the inverse Fourier transformation, we end up with Eq.~(\ref{eq:connection_E_j}).

\section{\label{app:solve_B=0} Plasma modes in an ellipse without a magnetic field}

To solve Eq.~(\ref{eq:potential_ell}) we use the method of separation of variables. For this purpose, we represent the scalar potential $\varphi$ as a product of functions depending only on the coordinate $\mu$ or $\nu$, i.e., let $\varphi(\mu,\nu)=M(\mu)N(\nu)$. Substituting this representation into Eq.~(\ref{eq:potential_ell}), we divide the equation by $M(\mu)N(\nu)$ and rearrange the terms to get the following form:
\begin{equation}
    \frac{N''(\nu)}{N(\nu)}-\frac{\omega^2f^2}{2 v^2}\cos 2\nu = -\frac{M''(\mu)}{M(\mu)}+\frac{\omega^2f^2}{2 v^2}\cosh 2\mu,
\end{equation}
where the stroke means the derivative. In the case, the boundary condition is $M'(\mu)\big||_{\mu=\mu_0} = 0$. Since this equality has to hold for any values of $\nu$ and $\mu$, the right and left sides of this equation are some number $-\alpha$. This statement leads us to explicit equations on the desired functions:
\begin{gather}
    N''(\nu)+(\alpha-2\xi \cos 2\nu) N(\nu)=0,\label{eq:N}\\
    M''(\mu)+(-\alpha+2\xi \cosh 2\mu) M(\mu)=0,\label{eq:M}
\end{gather}
where by $\xi$ we denote the value $\omega^2f^2/(2v)^2$. The resulting equations are called the Mathieu equation and the modified Mathieu equation, respectively \cite{Abramowitz1972}. 

Let us find the solutions of Eq.~(\ref{eq:N}). When we change the value of the coordinate $\nu$ from $0$ to $2\pi$ for a fixed $\mu$, we traverse around some ellipse and return to the initial point. Therefore, we should look for solutions satisfying the condition $N(0)=N(2\pi)$. From the theory of the Mathieu equation we know that for some values of the characteristic number $\alpha$ such solutions exist, and they are $2\pi$-periodic. These characteristic values depend on the value of $\xi$ and can be numbered. We will denote them as $\alpha_m(\xi)$, where $m$ is any integer. The characteristic numbers with $m\le 0$ and $m>0$ correspond to even $\text{ce}_m(\nu,\xi)$ and odd $\text{se}_{|m|}(\nu,\xi)$ solutions, respectively, which are called Mathieu functions of the first kind. The characteristic numbers satisfy a chain of strict inequalities $\alpha_{-|m|}(\xi)< \alpha_{|m|+1}(\xi)\le \alpha_{-|m|-1}(\xi)$. The equality is achieved only in the isotropic case, i.e., $\xi=0$, where $\alpha_0(0)=0$ and $\alpha_m(0)=\alpha_{-m}(0)=m^2$ ($m\neq 0$), and the Mathieu functions tend to $\cos m\nu$ and $\sin m\nu$ ($\text{ce}_0(\nu,0)=1/\sqrt{2}$).

We now turn to a discussion of Eq.~(\ref{eq:M}). Note that substituting $\mu\rightarrow i\mu$ leads us to the same kind of equation as in Eq.~(\ref{eq:N}). Then, given a choice of characteristic values as $\alpha_m(\xi)$, the solution to this equation will also be Mathieu functions. These functions are called modified Mathieu functions and are denoted as $\text{Ce}_m(\mu,\xi)=\text{ce}_m(i\mu,\xi)$ and $\text{Se}_m(\mu,\xi)=-i\,\text{se}_m(i\mu,\xi)$. In the absence of anisotropy $f\rightarrow 0$, these functions should reduce to Bessel functions of the first kind of order $|m|$, but this is not immediately apparent from Eq.~(\ref{eq:M}). To demonstrate this, we substitute the variable $\mu=\text{arcosh}(\eta/2\sqrt{\xi})$. Then the differential equation takes the form:
\begin{multline}
    \frac{1}{\sqrt{\eta^2-4\xi}}\frac{d}{d\eta}\left(\sqrt{\eta^2-4\xi}\frac{d}{d\eta}\right)M(\eta)\\
    -\frac{\alpha_m(\xi)+2\xi-\eta^2}{\eta^2-4\xi}M(\eta)=0.
\end{multline}
In the isotropic case $\xi\rightarrow 0$ and $\alpha_m(0)=m^2$, which leads to the Bessel equation. Given that at the center of the disk, i.e., at $\eta=0$, the potential must be finite, we obtain the solution in the form of the Bessel function $J_{|m|}(\eta)$ of the first order of $|m|$.

In the boundary condition (\ref{eq:potential_ell}) in the absence of a magnetic field, the derivative acts only on the function $M(\mu)$. Thus, $\mu_0$ has to be the zero of the derivative of the modified Mathieu functions, which have a countable number of zeros numbered by the integer $n=1,2,...$. However, it is more convenient to rewrite this condition through the derivative of the argument $\eta$, since in this case the limit of the isotropic system is immediately traceable. Denoting the zeros by $\eta_{m}^{(n)}(\xi)$ we come to the expression for the frequency Eqs.~(\ref{eq:frequency_B=0}), (\ref{eq:alpha}).

\section{\label{app:solve_Bneq0} Magnetoplasma modes in an ellipse}


The system of equations (\ref{eq:system_Bne0}), (\ref{eq:matrix_Bne0}) decomposes into two independent subsystems. To demonstrate this, we should use the decomposition of Mathieu functions into Fourier series \cite{Abramowitz1972}. For even functions:
\begin{gather}
    \text{ce}_{2n}(\nu,\xi) = \sum_{k=0}^{\infty} A_{k}^{(2n)} \cos 2k\nu,\\
    \text{ce}_{2n+1}(\nu,\xi) = \sum_{k=0}^{\infty} A_{k}^{(2n+1)} \cos (2k+1)\nu,
\end{gather}
and odd ones:
\begin{gather}
    \text{se}_{2n+1}(\nu,\xi) = \sum_{k=0}^{\infty} B_{k}^{(2n+1)} \sin (2k+1)\nu,\\
    \text{se}_{2n+2}(\nu,\xi) = \sum_{k=1}^{\infty} B_{k}^{(2n+2)} \sin 2k\nu,
\end{gather}
where the number $n=0,1,2,...$. For further discussion, the explicit form of the coefficients $A$ and $B$ is not important. The integral included in the matrix element $B_{m,m'}$ of (\ref{eq:matrix_Bne0}) is antisymmetric, i.e., it changes its sign when the indices $m$ and $m'$ change. Using Fourier series expansions, we can write that the matrix element $B_{m,m'}$ is zero if 1) $m$ is positive, $m'$ is non-negative and vice versa, 2) $m + m'$ is an odd number. Fixing now in the system (\ref{eq:system_Bne0}) an arbitrary even $m$ (odd) only coefficients with even $m'$ (odd) remain in the series. Thus, the first subsystem describes the interaction of modes with even numbers $m$, and the second -- modes with odd numbers $m$.

\bibliography{main}

\begin{thebibliography}{41}%
\makeatletter
\providecommand \@ifxundefined [1]{%
 \@ifx{#1\undefined}
}%
\providecommand \@ifnum [1]{%
 \ifnum #1\expandafter \@firstoftwo
 \else \expandafter \@secondoftwo
 \fi
}%
\providecommand \@ifx [1]{%
 \ifx #1\expandafter \@firstoftwo
 \else \expandafter \@secondoftwo
 \fi
}%
\providecommand \natexlab [1]{#1}%
\providecommand \enquote  [1]{``#1''}%
\providecommand \bibnamefont  [1]{#1}%
\providecommand \bibfnamefont [1]{#1}%
\providecommand \citenamefont [1]{#1}%
\providecommand \href@noop [0]{\@secondoftwo}%
\providecommand \href [0]{\begingroup \@sanitize@url \@href}%
\providecommand \@href[1]{\@@startlink{#1}\@@href}%
\providecommand \@@href[1]{\endgroup#1\@@endlink}%
\providecommand \@sanitize@url [0]{\catcode `\\12\catcode `\$12\catcode `\&12\catcode `\#12\catcode `\^12\catcode `\_12\catcode `\%12\relax}%
\providecommand \@@startlink[1]{}%
\providecommand \@@endlink[0]{}%
\providecommand \url  [0]{\begingroup\@sanitize@url \@url }%
\providecommand \@url [1]{\endgroup\@href {#1}{\urlprefix }}%
\providecommand \urlprefix  [0]{URL }%
\providecommand \Eprint [0]{\href }%
\providecommand \doibase [0]{https://doi.org/}%
\providecommand \selectlanguage [0]{\@gobble}%
\providecommand \bibinfo  [0]{\@secondoftwo}%
\providecommand \bibfield  [0]{\@secondoftwo}%
\providecommand \translation [1]{[#1]}%
\providecommand \BibitemOpen [0]{}%
\providecommand \bibitemStop [0]{}%
\providecommand \bibitemNoStop [0]{.\EOS\space}%
\providecommand \EOS [0]{\spacefactor3000\relax}%
\providecommand \BibitemShut  [1]{\csname bibitem#1\endcsname}%
\let\auto@bib@innerbib\@empty
\bibitem [{\citenamefont {Shur}(2020)}]{Shur2020}%
  \BibitemOpen
  \bibfield  {author} {\bibinfo {author} {\bibfnamefont {M.~S.}\ \bibnamefont {Shur}},\ }\bibfield  {title} {\bibinfo {title} {Terahertz plasmonic technology},\ }\href {https://doi.org/10.1109/JSEN.2020.3022809} {\bibfield  {journal} {\bibinfo  {journal} {IEEE Sens. J.}\ }\textbf {\bibinfo {volume} {21}},\ \bibinfo {pages} {12752} (\bibinfo {year} {2020})}\BibitemShut {NoStop}%
\bibitem [{\citenamefont {Elbanna}\ \emph {et~al.}(2023)\citenamefont {Elbanna}, \citenamefont {Jiang}, \citenamefont {Fu}, \citenamefont {Zhu}, \citenamefont {Liu}, \citenamefont {Zhao}, \citenamefont {Liu}, \citenamefont {Lai}, \citenamefont {Chua}, \citenamefont {Pan} \emph {et~al.}}]{Elbanna2023}%
  \BibitemOpen
  \bibfield  {author} {\bibinfo {author} {\bibfnamefont {A.}~\bibnamefont {Elbanna}}, \bibinfo {author} {\bibfnamefont {H.}~\bibnamefont {Jiang}}, \bibinfo {author} {\bibfnamefont {Q.}~\bibnamefont {Fu}}, \bibinfo {author} {\bibfnamefont {J.-F.}\ \bibnamefont {Zhu}}, \bibinfo {author} {\bibfnamefont {Y.}~\bibnamefont {Liu}}, \bibinfo {author} {\bibfnamefont {M.}~\bibnamefont {Zhao}}, \bibinfo {author} {\bibfnamefont {D.}~\bibnamefont {Liu}}, \bibinfo {author} {\bibfnamefont {S.}~\bibnamefont {Lai}}, \bibinfo {author} {\bibfnamefont {X.~W.}\ \bibnamefont {Chua}}, \bibinfo {author} {\bibfnamefont {J.}~\bibnamefont {Pan}}, \emph {et~al.},\ }\bibfield  {title} {\bibinfo {title} {2d material infrared photonics and plasmonics},\ }\href {https://doi.org/10.1021/acsnano.2c10705} {\bibfield  {journal} {\bibinfo  {journal} {ACS Nano}\ }\textbf {\bibinfo {volume} {17}},\ \bibinfo {pages} {4134} (\bibinfo {year} {2023})}\BibitemShut {NoStop}%
\bibitem [{\citenamefont {Grimes}\ and\ \citenamefont {Adams}(1976)}]{Grimes1976}%
  \BibitemOpen
  \bibfield  {author} {\bibinfo {author} {\bibfnamefont {C.}~\bibnamefont {Grimes}}\ and\ \bibinfo {author} {\bibfnamefont {G.}~\bibnamefont {Adams}},\ }\bibfield  {title} {\bibinfo {title} {Observation of two-dimensional plasmons and electron-ripplon scattering in a sheet of electrons on liquid helium},\ }\href {https://doi.org/10.1103/PhysRevLett.36.145} {\bibfield  {journal} {\bibinfo  {journal} {Phys. Rev. Lett.}\ }\textbf {\bibinfo {volume} {36}},\ \bibinfo {pages} {145} (\bibinfo {year} {1976})}\BibitemShut {NoStop}%
\bibitem [{\citenamefont {Allen~Jr}\ \emph {et~al.}(1977)\citenamefont {Allen~Jr}, \citenamefont {Tsui},\ and\ \citenamefont {Logan}}]{Allen1977}%
  \BibitemOpen
  \bibfield  {author} {\bibinfo {author} {\bibfnamefont {S.}~\bibnamefont {Allen~Jr}}, \bibinfo {author} {\bibfnamefont {D.}~\bibnamefont {Tsui}},\ and\ \bibinfo {author} {\bibfnamefont {R.}~\bibnamefont {Logan}},\ }\bibfield  {title} {\bibinfo {title} {Observation of the two-dimensional plasmon in silicon inversion layers},\ }\href {https://doi.org/10.1103/PhysRevLett.38.980} {\bibfield  {journal} {\bibinfo  {journal} {Phys. Rev. Lett.}\ }\textbf {\bibinfo {volume} {38}},\ \bibinfo {pages} {980} (\bibinfo {year} {1977})}\BibitemShut {NoStop}%
\bibitem [{\citenamefont {Li}\ \emph {et~al.}(2014)\citenamefont {Li}, \citenamefont {Yu}, \citenamefont {Ye}, \citenamefont {Ge}, \citenamefont {Ou}, \citenamefont {Wu}, \citenamefont {Feng}, \citenamefont {Chen},\ and\ \citenamefont {Zhang}}]{Li2014}%
  \BibitemOpen
  \bibfield  {author} {\bibinfo {author} {\bibfnamefont {L.}~\bibnamefont {Li}}, \bibinfo {author} {\bibfnamefont {Y.}~\bibnamefont {Yu}}, \bibinfo {author} {\bibfnamefont {G.~J.}\ \bibnamefont {Ye}}, \bibinfo {author} {\bibfnamefont {Q.}~\bibnamefont {Ge}}, \bibinfo {author} {\bibfnamefont {X.}~\bibnamefont {Ou}}, \bibinfo {author} {\bibfnamefont {H.}~\bibnamefont {Wu}}, \bibinfo {author} {\bibfnamefont {D.}~\bibnamefont {Feng}}, \bibinfo {author} {\bibfnamefont {X.~H.}\ \bibnamefont {Chen}},\ and\ \bibinfo {author} {\bibfnamefont {Y.}~\bibnamefont {Zhang}},\ }\bibfield  {title} {\bibinfo {title} {Black phosphorus field-effect transistors},\ }\href {https://doi.org/10.1038/nnano.2014.35} {\bibfield  {journal} {\bibinfo  {journal} {Nat. Nanotech.}\ }\textbf {\bibinfo {volume} {9}},\ \bibinfo {pages} {372} (\bibinfo {year} {2014})}\BibitemShut {NoStop}%
\bibitem [{\citenamefont {Rodin}\ \emph {et~al.}(2014)\citenamefont {Rodin}, \citenamefont {Carvalho},\ and\ \citenamefont {Castro~Neto}}]{Rodin2014}%
  \BibitemOpen
  \bibfield  {author} {\bibinfo {author} {\bibfnamefont {A.~S.}\ \bibnamefont {Rodin}}, \bibinfo {author} {\bibfnamefont {A.}~\bibnamefont {Carvalho}},\ and\ \bibinfo {author} {\bibfnamefont {A.~H.}\ \bibnamefont {Castro~Neto}},\ }\bibfield  {title} {\bibinfo {title} {Strain-induced gap modification in black phosphorus},\ }\href {https://doi.org/10.1103/PhysRevLett.112.176801} {\bibfield  {journal} {\bibinfo  {journal} {Phys. Rev. Lett.}\ }\textbf {\bibinfo {volume} {112}},\ \bibinfo {pages} {176801} (\bibinfo {year} {2014})}\BibitemShut {NoStop}%
\bibitem [{\citenamefont {Qiao}\ \emph {et~al.}(2014)\citenamefont {Qiao}, \citenamefont {Kong}, \citenamefont {Hu}, \citenamefont {Yang},\ and\ \citenamefont {Ji}}]{Qiao2014}%
  \BibitemOpen
  \bibfield  {author} {\bibinfo {author} {\bibfnamefont {J.}~\bibnamefont {Qiao}}, \bibinfo {author} {\bibfnamefont {X.}~\bibnamefont {Kong}}, \bibinfo {author} {\bibfnamefont {Z.-X.}\ \bibnamefont {Hu}}, \bibinfo {author} {\bibfnamefont {F.}~\bibnamefont {Yang}},\ and\ \bibinfo {author} {\bibfnamefont {W.}~\bibnamefont {Ji}},\ }\bibfield  {title} {\bibinfo {title} {High-mobility transport anisotropy and linear dichroism in few-layer black phosphorus},\ }\href {https://doi.org/10.1038/ncomms5475} {\bibfield  {journal} {\bibinfo  {journal} {Nat. Commun.}\ }\textbf {\bibinfo {volume} {5}},\ \bibinfo {pages} {4475} (\bibinfo {year} {2014})}\BibitemShut {NoStop}%
\bibitem [{\citenamefont {Low}\ \emph {et~al.}(2014)\citenamefont {Low}, \citenamefont {Rold\'an}, \citenamefont {Wang}, \citenamefont {Xia}, \citenamefont {Avouris}, \citenamefont {Moreno},\ and\ \citenamefont {Guinea}}]{Low2014}%
  \BibitemOpen
  \bibfield  {author} {\bibinfo {author} {\bibfnamefont {T.}~\bibnamefont {Low}}, \bibinfo {author} {\bibfnamefont {R.}~\bibnamefont {Rold\'an}}, \bibinfo {author} {\bibfnamefont {H.}~\bibnamefont {Wang}}, \bibinfo {author} {\bibfnamefont {F.}~\bibnamefont {Xia}}, \bibinfo {author} {\bibfnamefont {P.}~\bibnamefont {Avouris}}, \bibinfo {author} {\bibfnamefont {L.~M.}\ \bibnamefont {Moreno}},\ and\ \bibinfo {author} {\bibfnamefont {F.}~\bibnamefont {Guinea}},\ }\bibfield  {title} {\bibinfo {title} {Plasmons and screening in monolayer and multilayer black phosphorus},\ }\href {https://doi.org/10.1103/PhysRevLett.113.106802} {\bibfield  {journal} {\bibinfo  {journal} {Phys. Rev. Lett.}\ }\textbf {\bibinfo {volume} {113}},\ \bibinfo {pages} {106802} (\bibinfo {year} {2014})}\BibitemShut {NoStop}%
\bibitem [{\citenamefont {Pogna}\ \emph {et~al.}(2024)\citenamefont {Pogna}, \citenamefont {Pistore}, \citenamefont {Viti}, \citenamefont {Li}, \citenamefont {Davies}, \citenamefont {Linfield},\ and\ \citenamefont {Vitiello}}]{Pogna2024}%
  \BibitemOpen
  \bibfield  {author} {\bibinfo {author} {\bibfnamefont {E.~A.}\ \bibnamefont {Pogna}}, \bibinfo {author} {\bibfnamefont {V.}~\bibnamefont {Pistore}}, \bibinfo {author} {\bibfnamefont {L.}~\bibnamefont {Viti}}, \bibinfo {author} {\bibfnamefont {L.}~\bibnamefont {Li}}, \bibinfo {author} {\bibfnamefont {A.~G.}\ \bibnamefont {Davies}}, \bibinfo {author} {\bibfnamefont {E.~H.}\ \bibnamefont {Linfield}},\ and\ \bibinfo {author} {\bibfnamefont {M.~S.}\ \bibnamefont {Vitiello}},\ }\bibfield  {title} {\bibinfo {title} {Near-field detection of gate-tunable anisotropic plasmon polaritons in black phosphorus at terahertz frequencies},\ }\href {https://doi.org/10.1038/s41467-024-45264-5} {\bibfield  {journal} {\bibinfo  {journal} {Nat. Commun.}\ }\textbf {\bibinfo {volume} {15}},\ \bibinfo {pages} {2373} (\bibinfo {year} {2024})}\BibitemShut {NoStop}%
\bibitem [{\citenamefont {Lin}\ \emph {et~al.}(2015)\citenamefont {Lin}, \citenamefont {Komsa}, \citenamefont {Yeh}, \citenamefont {Björkman}, \citenamefont {Liang}, \citenamefont {Ho}, \citenamefont {Huang}, \citenamefont {Chiu}, \citenamefont {Krasheninnikov},\ and\ \citenamefont {Suenaga}}]{Lin2015}%
  \BibitemOpen
  \bibfield  {author} {\bibinfo {author} {\bibfnamefont {Y.-C.}\ \bibnamefont {Lin}}, \bibinfo {author} {\bibfnamefont {H.-P.}\ \bibnamefont {Komsa}}, \bibinfo {author} {\bibfnamefont {C.-H.}\ \bibnamefont {Yeh}}, \bibinfo {author} {\bibfnamefont {T.}~\bibnamefont {Björkman}}, \bibinfo {author} {\bibfnamefont {Z.-Y.}\ \bibnamefont {Liang}}, \bibinfo {author} {\bibfnamefont {C.-H.}\ \bibnamefont {Ho}}, \bibinfo {author} {\bibfnamefont {Y.-S.}\ \bibnamefont {Huang}}, \bibinfo {author} {\bibfnamefont {P.-W.}\ \bibnamefont {Chiu}}, \bibinfo {author} {\bibfnamefont {A.~V.}\ \bibnamefont {Krasheninnikov}},\ and\ \bibinfo {author} {\bibfnamefont {K.}~\bibnamefont {Suenaga}},\ }\bibfield  {title} {\bibinfo {title} {Single-layer res2: Two-dimensional semiconductor with tunable in-plane anisotropy},\ }\href {https://doi.org/10.1021/acsnano.5b04851} {\bibfield  {journal} {\bibinfo  {journal} {ACS Nano}\ }\textbf {\bibinfo {volume} {9}},\ \bibinfo {pages} {11249} (\bibinfo {year} {2015})}\BibitemShut {NoStop}%
\bibitem [{\citenamefont {Chenet}\ \emph {et~al.}(2015)\citenamefont {Chenet}, \citenamefont {Aslan}, \citenamefont {Huang}, \citenamefont {Fan}, \citenamefont {van~der Zande}, \citenamefont {Heinz},\ and\ \citenamefont {Hone}}]{Chenet2015}%
  \BibitemOpen
  \bibfield  {author} {\bibinfo {author} {\bibfnamefont {D.~A.}\ \bibnamefont {Chenet}}, \bibinfo {author} {\bibfnamefont {B.}~\bibnamefont {Aslan}}, \bibinfo {author} {\bibfnamefont {P.~Y.}\ \bibnamefont {Huang}}, \bibinfo {author} {\bibfnamefont {C.}~\bibnamefont {Fan}}, \bibinfo {author} {\bibfnamefont {A.~M.}\ \bibnamefont {van~der Zande}}, \bibinfo {author} {\bibfnamefont {T.~F.}\ \bibnamefont {Heinz}},\ and\ \bibinfo {author} {\bibfnamefont {J.~C.}\ \bibnamefont {Hone}},\ }\bibfield  {title} {\bibinfo {title} {In-plane anisotropy in mono- and few-layer res2 probed by raman spectroscopy and scanning transmission electron microscopy},\ }\href {https://doi.org/10.1021/acs.nanolett.5b00910} {\bibfield  {journal} {\bibinfo  {journal} {Nano Lett.}\ }\textbf {\bibinfo {volume} {15}},\ \bibinfo {pages} {5667} (\bibinfo {year} {2015})}\BibitemShut {NoStop}%
\bibitem [{\citenamefont {Dai}\ and\ \citenamefont {Zeng}(2015)}]{Dai2015}%
  \BibitemOpen
  \bibfield  {author} {\bibinfo {author} {\bibfnamefont {J.}~\bibnamefont {Dai}}\ and\ \bibinfo {author} {\bibfnamefont {X.~C.}\ \bibnamefont {Zeng}},\ }\bibfield  {title} {\bibinfo {title} {Titanium trisulfide monolayer: Theoretical prediction of a new direct-gap semiconductor with high and anisotropic carrier mobility},\ }\href {https://doi.org/https://doi.org/10.1002/ange.201502107} {\bibfield  {journal} {\bibinfo  {journal} {Angew. Chem.}\ }\textbf {\bibinfo {volume} {127}},\ \bibinfo {pages} {7682} (\bibinfo {year} {2015})}\BibitemShut {NoStop}%
\bibitem [{\citenamefont {Saeed}\ \emph {et~al.}(2017)\citenamefont {Saeed}, \citenamefont {Kachmar},\ and\ \citenamefont {Carignano}}]{Saeed2017}%
  \BibitemOpen
  \bibfield  {author} {\bibinfo {author} {\bibfnamefont {Y.}~\bibnamefont {Saeed}}, \bibinfo {author} {\bibfnamefont {A.}~\bibnamefont {Kachmar}},\ and\ \bibinfo {author} {\bibfnamefont {M.~A.}\ \bibnamefont {Carignano}},\ }\bibfield  {title} {\bibinfo {title} {First-principles study of the transport properties in bulk and monolayer mx3 (m = ti, zr, hf and x = s, se) compounds},\ }\href {https://doi.org/10.1021/acs.jpcc.6b08067} {\bibfield  {journal} {\bibinfo  {journal} {J. Phys. Chem. C}\ }\textbf {\bibinfo {volume} {121}},\ \bibinfo {pages} {1399} (\bibinfo {year} {2017})}\BibitemShut {NoStop}%
\bibitem [{\citenamefont {Shayegan}\ \emph {et~al.}(2006)\citenamefont {Shayegan}, \citenamefont {De~Poortere}, \citenamefont {Gunawan}, \citenamefont {Shkolnikov}, \citenamefont {Tutuc},\ and\ \citenamefont {Vakili}}]{Shayegan2006}%
  \BibitemOpen
  \bibfield  {author} {\bibinfo {author} {\bibfnamefont {M.}~\bibnamefont {Shayegan}}, \bibinfo {author} {\bibfnamefont {E.~P.}\ \bibnamefont {De~Poortere}}, \bibinfo {author} {\bibfnamefont {O.}~\bibnamefont {Gunawan}}, \bibinfo {author} {\bibfnamefont {Y.~P.}\ \bibnamefont {Shkolnikov}}, \bibinfo {author} {\bibfnamefont {E.}~\bibnamefont {Tutuc}},\ and\ \bibinfo {author} {\bibfnamefont {K.}~\bibnamefont {Vakili}},\ }\bibfield  {title} {\bibinfo {title} {Two-dimensional electrons occupying multiple valleys in alas},\ }\href {https://doi.org/10.1002/pssb.200642212} {\bibfield  {journal} {\bibinfo  {journal} {Phys. Status Solidi (b)}\ }\textbf {\bibinfo {volume} {243}},\ \bibinfo {pages} {3629} (\bibinfo {year} {2006})}\BibitemShut {NoStop}%
\bibitem [{\citenamefont {Khisameeva}\ \emph {et~al.}(2018)\citenamefont {Khisameeva}, \citenamefont {Shchepetilnikov}, \citenamefont {Muravev}, \citenamefont {Gubarev}, \citenamefont {Frolov}, \citenamefont {Nefyodov}, \citenamefont {Kukushkin}, \citenamefont {Reichl}, \citenamefont {Tiemann}, \citenamefont {Dietsche},\ and\ \citenamefont {Wegscheider}}]{Khisameeva2018}%
  \BibitemOpen
  \bibfield  {author} {\bibinfo {author} {\bibfnamefont {A.~R.}\ \bibnamefont {Khisameeva}}, \bibinfo {author} {\bibfnamefont {A.~V.}\ \bibnamefont {Shchepetilnikov}}, \bibinfo {author} {\bibfnamefont {V.~M.}\ \bibnamefont {Muravev}}, \bibinfo {author} {\bibfnamefont {S.~I.}\ \bibnamefont {Gubarev}}, \bibinfo {author} {\bibfnamefont {D.~D.}\ \bibnamefont {Frolov}}, \bibinfo {author} {\bibfnamefont {Y.~A.}\ \bibnamefont {Nefyodov}}, \bibinfo {author} {\bibfnamefont {I.~V.}\ \bibnamefont {Kukushkin}}, \bibinfo {author} {\bibfnamefont {C.}~\bibnamefont {Reichl}}, \bibinfo {author} {\bibfnamefont {L.}~\bibnamefont {Tiemann}}, \bibinfo {author} {\bibfnamefont {W.}~\bibnamefont {Dietsche}},\ and\ \bibinfo {author} {\bibfnamefont {W.}~\bibnamefont {Wegscheider}},\ }\bibfield  {title} {\bibinfo {title} {Direct observation of a $\mathrm{\ensuremath{\Gamma}}\ensuremath{-}x$ energy spectrum transition in narrow alas quantum wells},\ }\href {https://doi.org/10.1103/PhysRevB.97.115308} {\bibfield  {journal} {\bibinfo
  {journal} {Phys. Rev. B}\ }\textbf {\bibinfo {volume} {97}},\ \bibinfo {pages} {115308} (\bibinfo {year} {2018})}\BibitemShut {NoStop}%
\bibitem [{\citenamefont {Khisameeva}\ \emph {et~al.}(2020)\citenamefont {Khisameeva}, \citenamefont {Muravev},\ and\ \citenamefont {Kukushkin}}]{Khisameeva2020}%
  \BibitemOpen
  \bibfield  {author} {\bibinfo {author} {\bibfnamefont {A.~R.}\ \bibnamefont {Khisameeva}}, \bibinfo {author} {\bibfnamefont {V.~M.}\ \bibnamefont {Muravev}},\ and\ \bibinfo {author} {\bibfnamefont {I.~V.}\ \bibnamefont {Kukushkin}},\ }\bibfield  {title} {\bibinfo {title} {Piezoplasmonics: Strain-induced tunability of plasmon resonance in alas quantum wells},\ }\href {https://doi.org/10.1063/5.0012496} {\bibfield  {journal} {\bibinfo  {journal} {Appl. Phys. Lett.}\ }\textbf {\bibinfo {volume} {117}},\ \bibinfo {pages} {093102} (\bibinfo {year} {2020})}\BibitemShut {NoStop}%
\bibitem [{\citenamefont {Ahn}\ and\ \citenamefont {Das~Sarma}(2021)}]{Ahn2021}%
  \BibitemOpen
  \bibfield  {author} {\bibinfo {author} {\bibfnamefont {S.}~\bibnamefont {Ahn}}\ and\ \bibinfo {author} {\bibfnamefont {S.}~\bibnamefont {Das~Sarma}},\ }\bibfield  {title} {\bibinfo {title} {Theory of anisotropic plasmons},\ }\href {https://doi.org/10.1103/PhysRevB.103.L041303} {\bibfield  {journal} {\bibinfo  {journal} {Phys. Rev. B}\ }\textbf {\bibinfo {volume} {103}},\ \bibinfo {pages} {L041303} (\bibinfo {year} {2021})}\BibitemShut {NoStop}%
\bibitem [{\citenamefont {Sokolik}\ \emph {et~al.}(2021)\citenamefont {Sokolik}, \citenamefont {Kotov},\ and\ \citenamefont {Lozovik}}]{Sokolik2021}%
  \BibitemOpen
  \bibfield  {author} {\bibinfo {author} {\bibfnamefont {A.~A.}\ \bibnamefont {Sokolik}}, \bibinfo {author} {\bibfnamefont {O.~V.}\ \bibnamefont {Kotov}},\ and\ \bibinfo {author} {\bibfnamefont {Y.~E.}\ \bibnamefont {Lozovik}},\ }\bibfield  {title} {\bibinfo {title} {Plasmonic modes at inclined edges of anisotropic two-dimensional materials},\ }\href {https://doi.org/10.1103/PhysRevB.103.155402} {\bibfield  {journal} {\bibinfo  {journal} {Phys. Rev. B}\ }\textbf {\bibinfo {volume} {103}},\ \bibinfo {pages} {155402} (\bibinfo {year} {2021})}\BibitemShut {NoStop}%
\bibitem [{\citenamefont {Mi\ifmmode \check{s}\else \v{s}\fi{}kovi\ifmmode~\acute{c}\else \'{c}\fi{}}\ and\ \citenamefont {Moshayedi}(2023)}]{Miskovic2023}%
  \BibitemOpen
  \bibfield  {author} {\bibinfo {author} {\bibfnamefont {Z.~L.}\ \bibnamefont {Mi\ifmmode \check{s}\else \v{s}\fi{}kovi\ifmmode~\acute{c}\else \'{c}\fi{}}}\ and\ \bibinfo {author} {\bibfnamefont {M.}~\bibnamefont {Moshayedi}},\ }\bibfield  {title} {\bibinfo {title} {Plasmon wake in anisotropic two-dimensional materials},\ }\href {https://doi.org/10.1103/PhysRevResearch.5.033133} {\bibfield  {journal} {\bibinfo  {journal} {Phys. Rev. Res.}\ }\textbf {\bibinfo {volume} {5}},\ \bibinfo {pages} {033133} (\bibinfo {year} {2023})}\BibitemShut {NoStop}%
\bibitem [{\citenamefont {Nemilentsau}\ \emph {et~al.}(2016)\citenamefont {Nemilentsau}, \citenamefont {Low},\ and\ \citenamefont {Hanson}}]{Nemilentsau2016}%
  \BibitemOpen
  \bibfield  {author} {\bibinfo {author} {\bibfnamefont {A.}~\bibnamefont {Nemilentsau}}, \bibinfo {author} {\bibfnamefont {T.}~\bibnamefont {Low}},\ and\ \bibinfo {author} {\bibfnamefont {G.}~\bibnamefont {Hanson}},\ }\bibfield  {title} {\bibinfo {title} {Anisotropic 2d materials for tunable hyperbolic plasmonics},\ }\href {https://doi.org/10.1103/PhysRevLett.116.066804} {\bibfield  {journal} {\bibinfo  {journal} {Phys. Rev. Lett.}\ }\textbf {\bibinfo {volume} {116}},\ \bibinfo {pages} {066804} (\bibinfo {year} {2016})}\BibitemShut {NoStop}%
\bibitem [{\citenamefont {Ma}\ \emph {et~al.}(2018)\citenamefont {Ma}, \citenamefont {Alonso-Gonz{\'a}lez}, \citenamefont {Li}, \citenamefont {Nikitin}, \citenamefont {Yuan}, \citenamefont {Mart{\'\i}n-S{\'a}nchez}, \citenamefont {Taboada-Guti{\'e}rrez}, \citenamefont {Amenabar}, \citenamefont {Li}, \citenamefont {V{\'e}lez} \emph {et~al.}}]{Ma2018}%
  \BibitemOpen
  \bibfield  {author} {\bibinfo {author} {\bibfnamefont {W.}~\bibnamefont {Ma}}, \bibinfo {author} {\bibfnamefont {P.}~\bibnamefont {Alonso-Gonz{\'a}lez}}, \bibinfo {author} {\bibfnamefont {S.}~\bibnamefont {Li}}, \bibinfo {author} {\bibfnamefont {A.~Y.}\ \bibnamefont {Nikitin}}, \bibinfo {author} {\bibfnamefont {J.}~\bibnamefont {Yuan}}, \bibinfo {author} {\bibfnamefont {J.}~\bibnamefont {Mart{\'\i}n-S{\'a}nchez}}, \bibinfo {author} {\bibfnamefont {J.}~\bibnamefont {Taboada-Guti{\'e}rrez}}, \bibinfo {author} {\bibfnamefont {I.}~\bibnamefont {Amenabar}}, \bibinfo {author} {\bibfnamefont {P.}~\bibnamefont {Li}}, \bibinfo {author} {\bibfnamefont {S.}~\bibnamefont {V{\'e}lez}}, \emph {et~al.},\ }\bibfield  {title} {\bibinfo {title} {In-plane anisotropic and ultra-low-loss polaritons in a natural van der waals crystal},\ }\href {https://doi.org/10.1038/s41586-018-0618-9} {\bibfield  {journal} {\bibinfo  {journal} {Nature}\ }\textbf {\bibinfo {volume} {562}},\ \bibinfo {pages} {557} (\bibinfo {year}
  {2018})}\BibitemShut {NoStop}%
\bibitem [{\citenamefont {Margetis}\ \emph {et~al.}(2020)\citenamefont {Margetis}, \citenamefont {Maier}, \citenamefont {Stauber}, \citenamefont {Low},\ and\ \citenamefont {Luskin}}]{Margetis2020}%
  \BibitemOpen
  \bibfield  {author} {\bibinfo {author} {\bibfnamefont {D.}~\bibnamefont {Margetis}}, \bibinfo {author} {\bibfnamefont {M.}~\bibnamefont {Maier}}, \bibinfo {author} {\bibfnamefont {T.}~\bibnamefont {Stauber}}, \bibinfo {author} {\bibfnamefont {T.}~\bibnamefont {Low}},\ and\ \bibinfo {author} {\bibfnamefont {M.}~\bibnamefont {Luskin}},\ }\bibfield  {title} {\bibinfo {title} {Nonretarded edge plasmon-polaritons in anisotropic two-dimensional materials},\ }\href {https://doi.org/10.1088/1751-8121/ab5ff9} {\bibfield  {journal} {\bibinfo  {journal} {J. Phys. A: Math. Theor.}\ }\textbf {\bibinfo {volume} {53}},\ \bibinfo {pages} {055201} (\bibinfo {year} {2020})}\BibitemShut {NoStop}%
\bibitem [{\citenamefont {Jin}\ \emph {et~al.}(2016)\citenamefont {Jin}, \citenamefont {Lu}, \citenamefont {Wang}, \citenamefont {Fang}, \citenamefont {Joannopoulos}, \citenamefont {Solja{\v{c}}i{\'c}}, \citenamefont {Fu},\ and\ \citenamefont {Fang}}]{Jin2016}%
  \BibitemOpen
  \bibfield  {author} {\bibinfo {author} {\bibfnamefont {D.}~\bibnamefont {Jin}}, \bibinfo {author} {\bibfnamefont {L.}~\bibnamefont {Lu}}, \bibinfo {author} {\bibfnamefont {Z.}~\bibnamefont {Wang}}, \bibinfo {author} {\bibfnamefont {C.}~\bibnamefont {Fang}}, \bibinfo {author} {\bibfnamefont {J.~D.}\ \bibnamefont {Joannopoulos}}, \bibinfo {author} {\bibfnamefont {M.}~\bibnamefont {Solja{\v{c}}i{\'c}}}, \bibinfo {author} {\bibfnamefont {L.}~\bibnamefont {Fu}},\ and\ \bibinfo {author} {\bibfnamefont {N.~X.}\ \bibnamefont {Fang}},\ }\bibfield  {title} {\bibinfo {title} {Topological magnetoplasmon},\ }\href {https://doi.org/10.1038/ncomms13486} {\bibfield  {journal} {\bibinfo  {journal} {Nat. Commun.}\ }\textbf {\bibinfo {volume} {7}},\ \bibinfo {pages} {13486} (\bibinfo {year} {2016})}\BibitemShut {NoStop}%
\bibitem [{\citenamefont {Zagorodnev}\ \emph {et~al.}(2023)\citenamefont {Zagorodnev}, \citenamefont {Zabolotnykh}, \citenamefont {Rodionov},\ and\ \citenamefont {Volkov}}]{Zagorodnev2023}%
  \BibitemOpen
  \bibfield  {author} {\bibinfo {author} {\bibfnamefont {I.~V.}\ \bibnamefont {Zagorodnev}}, \bibinfo {author} {\bibfnamefont {A.~A.}\ \bibnamefont {Zabolotnykh}}, \bibinfo {author} {\bibfnamefont {D.~A.}\ \bibnamefont {Rodionov}},\ and\ \bibinfo {author} {\bibfnamefont {V.~A.}\ \bibnamefont {Volkov}},\ }\bibfield  {title} {\bibinfo {title} {Two-dimensional plasmons in laterally confined 2d electron systems},\ }\href {https://doi.org/10.3390/nano13060975} {\bibfield  {journal} {\bibinfo  {journal} {Nanomaterials}\ }\textbf {\bibinfo {volume} {13}},\ \bibinfo {pages} {975} (\bibinfo {year} {2023})}\BibitemShut {NoStop}%
\bibitem [{\citenamefont {Rodionov}\ and\ \citenamefont {Zagorodnev}(2023)}]{Rodionov2023}%
  \BibitemOpen
  \bibfield  {author} {\bibinfo {author} {\bibfnamefont {D.}~\bibnamefont {Rodionov}}\ and\ \bibinfo {author} {\bibfnamefont {I.}~\bibnamefont {Zagorodnev}},\ }\bibfield  {title} {\bibinfo {title} {Plasmons in a strip with an anisotropic two-dimensional electron gas fully screened by a metal gate},\ }\href {https://doi.org/10.1134/S0021364023601811} {\bibfield  {journal} {\bibinfo  {journal} {JETP Lett.}\ }\textbf {\bibinfo {volume} {118}},\ \bibinfo {pages} {100} (\bibinfo {year} {2023})}\BibitemShut {NoStop}%
\bibitem [{\citenamefont {Muravev}\ \emph {et~al.}(2007)\citenamefont {Muravev}, \citenamefont {Jiang}, \citenamefont {Kukushkin}, \citenamefont {Smet}, \citenamefont {Umansky},\ and\ \citenamefont {von Klitzing}}]{Muravev2007}%
  \BibitemOpen
  \bibfield  {author} {\bibinfo {author} {\bibfnamefont {V.~M.}\ \bibnamefont {Muravev}}, \bibinfo {author} {\bibfnamefont {C.}~\bibnamefont {Jiang}}, \bibinfo {author} {\bibfnamefont {I.~V.}\ \bibnamefont {Kukushkin}}, \bibinfo {author} {\bibfnamefont {J.~H.}\ \bibnamefont {Smet}}, \bibinfo {author} {\bibfnamefont {V.}~\bibnamefont {Umansky}},\ and\ \bibinfo {author} {\bibfnamefont {K.}~\bibnamefont {von Klitzing}},\ }\bibfield  {title} {\bibinfo {title} {Spectra of magnetoplasma excitations in back-gate hall bar structures},\ }\href {https://doi.org/10.1103/PhysRevB.75.193307} {\bibfield  {journal} {\bibinfo  {journal} {Phys. Rev. B}\ }\textbf {\bibinfo {volume} {75}},\ \bibinfo {pages} {193307} (\bibinfo {year} {2007})}\BibitemShut {NoStop}%
\bibitem [{\citenamefont {Bandurin}\ \emph {et~al.}(2018)\citenamefont {Bandurin}, \citenamefont {Svintsov}, \citenamefont {Gayduchenko}, \citenamefont {Xu}, \citenamefont {Principi}, \citenamefont {Moskotin}, \citenamefont {Tretyakov}, \citenamefont {Yagodkin}, \citenamefont {Zhukov}, \citenamefont {Taniguchi} \emph {et~al.}}]{Bandurin2018}%
  \BibitemOpen
  \bibfield  {author} {\bibinfo {author} {\bibfnamefont {D.~A.}\ \bibnamefont {Bandurin}}, \bibinfo {author} {\bibfnamefont {D.}~\bibnamefont {Svintsov}}, \bibinfo {author} {\bibfnamefont {I.}~\bibnamefont {Gayduchenko}}, \bibinfo {author} {\bibfnamefont {S.~G.}\ \bibnamefont {Xu}}, \bibinfo {author} {\bibfnamefont {A.}~\bibnamefont {Principi}}, \bibinfo {author} {\bibfnamefont {M.}~\bibnamefont {Moskotin}}, \bibinfo {author} {\bibfnamefont {I.}~\bibnamefont {Tretyakov}}, \bibinfo {author} {\bibfnamefont {D.}~\bibnamefont {Yagodkin}}, \bibinfo {author} {\bibfnamefont {S.}~\bibnamefont {Zhukov}}, \bibinfo {author} {\bibfnamefont {T.}~\bibnamefont {Taniguchi}}, \emph {et~al.},\ }\bibfield  {title} {\bibinfo {title} {Resonant terahertz detection using graphene plasmons},\ }\href {https://doi.org/10.1038/s41467-018-07848-w} {\bibfield  {journal} {\bibinfo  {journal} {Nat. Commun.}\ }\textbf {\bibinfo {volume} {9}},\ \bibinfo {pages} {5392} (\bibinfo {year} {2018})}\BibitemShut {NoStop}%
\bibitem [{\citenamefont {Alcaraz~Iranzo}\ \emph {et~al.}(2018)\citenamefont {Alcaraz~Iranzo}, \citenamefont {Nanot}, \citenamefont {Dias}, \citenamefont {Epstein}, \citenamefont {Peng}, \citenamefont {Efetov}, \citenamefont {Lundeberg}, \citenamefont {Parret}, \citenamefont {Osmond}, \citenamefont {Hong} \emph {et~al.}}]{Alcaraz2018}%
  \BibitemOpen
  \bibfield  {author} {\bibinfo {author} {\bibfnamefont {D.}~\bibnamefont {Alcaraz~Iranzo}}, \bibinfo {author} {\bibfnamefont {S.}~\bibnamefont {Nanot}}, \bibinfo {author} {\bibfnamefont {E.~J.}\ \bibnamefont {Dias}}, \bibinfo {author} {\bibfnamefont {I.}~\bibnamefont {Epstein}}, \bibinfo {author} {\bibfnamefont {C.}~\bibnamefont {Peng}}, \bibinfo {author} {\bibfnamefont {D.~K.}\ \bibnamefont {Efetov}}, \bibinfo {author} {\bibfnamefont {M.~B.}\ \bibnamefont {Lundeberg}}, \bibinfo {author} {\bibfnamefont {R.}~\bibnamefont {Parret}}, \bibinfo {author} {\bibfnamefont {J.}~\bibnamefont {Osmond}}, \bibinfo {author} {\bibfnamefont {J.-Y.}\ \bibnamefont {Hong}}, \emph {et~al.},\ }\bibfield  {title} {\bibinfo {title} {Probing the ultimate plasmon confinement limits with a van der waals heterostructure},\ }\href {https://doi.org/10.1126/science.aar843} {\bibfield  {journal} {\bibinfo  {journal} {Science}\ }\textbf {\bibinfo {volume} {360}},\ \bibinfo {pages} {291} (\bibinfo {year} {2018})}\BibitemShut {NoStop}%
\bibitem [{\citenamefont {Bylinkin}\ \emph {et~al.}(2019)\citenamefont {Bylinkin}, \citenamefont {Titova}, \citenamefont {Mikheev}, \citenamefont {Zhukova}, \citenamefont {Zhukov}, \citenamefont {Belyanchikov}, \citenamefont {Kashchenko}, \citenamefont {Miakonkikh},\ and\ \citenamefont {Svintsov}}]{Bylinkin2019}%
  \BibitemOpen
  \bibfield  {author} {\bibinfo {author} {\bibfnamefont {A.}~\bibnamefont {Bylinkin}}, \bibinfo {author} {\bibfnamefont {E.}~\bibnamefont {Titova}}, \bibinfo {author} {\bibfnamefont {V.}~\bibnamefont {Mikheev}}, \bibinfo {author} {\bibfnamefont {E.}~\bibnamefont {Zhukova}}, \bibinfo {author} {\bibfnamefont {S.}~\bibnamefont {Zhukov}}, \bibinfo {author} {\bibfnamefont {M.}~\bibnamefont {Belyanchikov}}, \bibinfo {author} {\bibfnamefont {M.}~\bibnamefont {Kashchenko}}, \bibinfo {author} {\bibfnamefont {A.}~\bibnamefont {Miakonkikh}},\ and\ \bibinfo {author} {\bibfnamefont {D.}~\bibnamefont {Svintsov}},\ }\bibfield  {title} {\bibinfo {title} {Tight-binding terahertz plasmons in chemical-vapor-deposited graphene},\ }\href {https://doi.org/10.1103/PhysRevApplied.11.054017} {\bibfield  {journal} {\bibinfo  {journal} {Phys. Rev. Appl.}\ }\textbf {\bibinfo {volume} {11}},\ \bibinfo {pages} {054017} (\bibinfo {year} {2019})}\BibitemShut {NoStop}%
\bibitem [{\citenamefont {Mikhailov}\ and\ \citenamefont {Savostianova}(2005)}]{Mikhailov2005}%
  \BibitemOpen
  \bibfield  {author} {\bibinfo {author} {\bibfnamefont {S.~A.}\ \bibnamefont {Mikhailov}}\ and\ \bibinfo {author} {\bibfnamefont {N.~A.}\ \bibnamefont {Savostianova}},\ }\bibfield  {title} {\bibinfo {title} {Microwave response of a two-dimensional electron stripe},\ }\href {https://doi.org/10.1103/PhysRevB.71.035320} {\bibfield  {journal} {\bibinfo  {journal} {Phys. Rev. B}\ }\textbf {\bibinfo {volume} {71}},\ \bibinfo {pages} {035320} (\bibinfo {year} {2005})}\BibitemShut {NoStop}%
\bibitem [{\citenamefont {Fetter}(1986)}]{Fetter1986}%
  \BibitemOpen
  \bibfield  {author} {\bibinfo {author} {\bibfnamefont {A.~L.}\ \bibnamefont {Fetter}},\ }\bibfield  {title} {\bibinfo {title} {Magnetoplasmons in a two-dimensional electron fluid: Disk geometry},\ }\href {https://doi.org/10.1103/PhysRevB.33.5221} {\bibfield  {journal} {\bibinfo  {journal} {Phys. Rev. B}\ }\textbf {\bibinfo {volume} {33}},\ \bibinfo {pages} {5221} (\bibinfo {year} {1986})}\BibitemShut {NoStop}%
\bibitem [{\citenamefont {Volkov}\ and\ \citenamefont {Mikhailov}(1988)}]{Volkov1988}%
  \BibitemOpen
  \bibfield  {author} {\bibinfo {author} {\bibfnamefont {V.~A.}\ \bibnamefont {Volkov}}\ and\ \bibinfo {author} {\bibfnamefont {S.~A.}\ \bibnamefont {Mikhailov}},\ }\bibfield  {title} {\bibinfo {title} {Edge magnetoplasmons: low frequency weakly damped excitations in inhomogeneous two-dimensional electron systems},\ }\href {http://jetp.ras.ru/cgi-bin/dn/e_067_08_1639.pdf} {\bibfield  {journal} {\bibinfo  {journal} {Sov. Phys. JETP}\ }\textbf {\bibinfo {volume} {67}},\ \bibinfo {pages} {1639} (\bibinfo {year} {1988})}\BibitemShut {NoStop}%
\bibitem [{\citenamefont {Gabov}(1978)}]{Gabov1976}%
  \BibitemOpen
  \bibfield  {author} {\bibinfo {author} {\bibfnamefont {S.}~\bibnamefont {Gabov}},\ }\bibfield  {title} {\bibinfo {title} {Green's function of the laplace operator in a two-dimensional problem with directional derivative with constant coefficients},\ }\href {https://doi.org/10.1016/0041-5553(78)90174-X} {\bibfield  {journal} {\bibinfo  {journal} {USSR Comput. Math. Math. Phys.}\ }\textbf {\bibinfo {volume} {18}},\ \bibinfo {pages} {132} (\bibinfo {year} {1978})}\BibitemShut {NoStop}%
\bibitem [{\citenamefont {Volkov}\ and\ \citenamefont {Mikhailov}(1985)}]{Volkov1985}%
  \BibitemOpen
  \bibfield  {author} {\bibinfo {author} {\bibfnamefont {V.~A.}\ \bibnamefont {Volkov}}\ and\ \bibinfo {author} {\bibfnamefont {S.~A.}\ \bibnamefont {Mikhailov}},\ }\bibfield  {title} {\bibinfo {title} {Theory of edge magnetoplasmons in a two-dimensional electron gas},\ }\href {http://jetpletters.ru/ps/1439/article_21888.pdf} {\bibfield  {journal} {\bibinfo  {journal} {JETP Lett.}\ }\textbf {\bibinfo {volume} {42}},\ \bibinfo {pages} {556} (\bibinfo {year} {1985})}\BibitemShut {NoStop}%
\bibitem [{\citenamefont {Mast}\ \emph {et~al.}(1985)\citenamefont {Mast}, \citenamefont {Dahm},\ and\ \citenamefont {Fetter}}]{Mast1985}%
  \BibitemOpen
  \bibfield  {author} {\bibinfo {author} {\bibfnamefont {D.~B.}\ \bibnamefont {Mast}}, \bibinfo {author} {\bibfnamefont {A.~J.}\ \bibnamefont {Dahm}},\ and\ \bibinfo {author} {\bibfnamefont {A.~L.}\ \bibnamefont {Fetter}},\ }\bibfield  {title} {\bibinfo {title} {Observation of bulk and edge magnetoplasmons in a two-dimensional electron fluid},\ }\href {https://doi.org/10.1103/PhysRevLett.54.1706} {\bibfield  {journal} {\bibinfo  {journal} {Phys. Rev. Lett.}\ }\textbf {\bibinfo {volume} {54}},\ \bibinfo {pages} {1706} (\bibinfo {year} {1985})}\BibitemShut {NoStop}%
\bibitem [{\citenamefont {Zabolotnykh}\ and\ \citenamefont {Volkov}(2021)}]{Zabolotnykh2021}%
  \BibitemOpen
  \bibfield  {author} {\bibinfo {author} {\bibfnamefont {A.~A.}\ \bibnamefont {Zabolotnykh}}\ and\ \bibinfo {author} {\bibfnamefont {V.~A.}\ \bibnamefont {Volkov}},\ }\bibfield  {title} {\bibinfo {title} {Electrically controllable cyclotron resonance},\ }\href {https://doi.org/10.1103/PhysRevB.103.125301} {\bibfield  {journal} {\bibinfo  {journal} {Phys. Rev. B}\ }\textbf {\bibinfo {volume} {103}},\ \bibinfo {pages} {125301} (\bibinfo {year} {2021})}\BibitemShut {NoStop}%
\bibitem [{\citenamefont {Rodionov}\ and\ \citenamefont {Zagorodnev}(2024)}]{Rodionov2024}%
  \BibitemOpen
  \bibfield  {author} {\bibinfo {author} {\bibfnamefont {D.~A.}\ \bibnamefont {Rodionov}}\ and\ \bibinfo {author} {\bibfnamefont {I.~V.}\ \bibnamefont {Zagorodnev}},\ }\bibfield  {title} {\bibinfo {title} {Fully screened two-dimensional magnetoplasmons and rotational gravity shallow water waves in a rectangle},\ }\href {https://doi.org/10.1103/PhysRevB.109.L241402} {\bibfield  {journal} {\bibinfo  {journal} {Phys. Rev. B}\ }\textbf {\bibinfo {volume} {109}},\ \bibinfo {pages} {L241402} (\bibinfo {year} {2024})}\BibitemShut {NoStop}%
\bibitem [{\citenamefont {Zarezin}\ \emph {et~al.}(2023)\citenamefont {Zarezin}, \citenamefont {Mylnikov}, \citenamefont {Petrov}, \citenamefont {Svintsov}, \citenamefont {Gusikhin}, \citenamefont {Kukushkin},\ and\ \citenamefont {Muravev}}]{Zarezin2023}%
  \BibitemOpen
  \bibfield  {author} {\bibinfo {author} {\bibfnamefont {A.~M.}\ \bibnamefont {Zarezin}}, \bibinfo {author} {\bibfnamefont {D.}~\bibnamefont {Mylnikov}}, \bibinfo {author} {\bibfnamefont {A.~S.}\ \bibnamefont {Petrov}}, \bibinfo {author} {\bibfnamefont {D.}~\bibnamefont {Svintsov}}, \bibinfo {author} {\bibfnamefont {P.~A.}\ \bibnamefont {Gusikhin}}, \bibinfo {author} {\bibfnamefont {I.~V.}\ \bibnamefont {Kukushkin}},\ and\ \bibinfo {author} {\bibfnamefont {V.~M.}\ \bibnamefont {Muravev}},\ }\bibfield  {title} {\bibinfo {title} {Plasmons in a square of two-dimensional electrons},\ }\href {https://doi.org/10.1103/PhysRevB.107.075414} {\bibfield  {journal} {\bibinfo  {journal} {Phys. Rev. B}\ }\textbf {\bibinfo {volume} {107}},\ \bibinfo {pages} {075414} (\bibinfo {year} {2023})}\BibitemShut {NoStop}%
\bibitem [{\citenamefont {Gusikhin}\ \emph {et~al.}(2018)\citenamefont {Gusikhin}, \citenamefont {Muravev}, \citenamefont {Zagitova},\ and\ \citenamefont {Kukushkin}}]{Gusikhin2018}%
  \BibitemOpen
  \bibfield  {author} {\bibinfo {author} {\bibfnamefont {P.~A.}\ \bibnamefont {Gusikhin}}, \bibinfo {author} {\bibfnamefont {V.~M.}\ \bibnamefont {Muravev}}, \bibinfo {author} {\bibfnamefont {A.~A.}\ \bibnamefont {Zagitova}},\ and\ \bibinfo {author} {\bibfnamefont {I.~V.}\ \bibnamefont {Kukushkin}},\ }\bibfield  {title} {\bibinfo {title} {Drastic reduction of plasmon damping in two-dimensional electron disks},\ }\href {https://doi.org/10.1103/PhysRevLett.121.176804} {\bibfield  {journal} {\bibinfo  {journal} {Phys. Rev. Lett.}\ }\textbf {\bibinfo {volume} {121}},\ \bibinfo {pages} {176804} (\bibinfo {year} {2018})}\BibitemShut {NoStop}%
\bibitem [{\citenamefont {Chen}\ \emph {et~al.}(2023)\citenamefont {Chen}, \citenamefont {Yin}, \citenamefont {Zhang}, \citenamefont {Yuan}, \citenamefont {Xu}, \citenamefont {Zhang}, \citenamefont {Chen}, \citenamefont {Zhang}, \citenamefont {Li}, \citenamefont {Wang} \emph {et~al.}}]{Chen2023}%
  \BibitemOpen
  \bibfield  {author} {\bibinfo {author} {\bibfnamefont {C.}~\bibnamefont {Chen}}, \bibinfo {author} {\bibfnamefont {Y.}~\bibnamefont {Yin}}, \bibinfo {author} {\bibfnamefont {R.}~\bibnamefont {Zhang}}, \bibinfo {author} {\bibfnamefont {Q.}~\bibnamefont {Yuan}}, \bibinfo {author} {\bibfnamefont {Y.}~\bibnamefont {Xu}}, \bibinfo {author} {\bibfnamefont {Y.}~\bibnamefont {Zhang}}, \bibinfo {author} {\bibfnamefont {J.}~\bibnamefont {Chen}}, \bibinfo {author} {\bibfnamefont {Y.}~\bibnamefont {Zhang}}, \bibinfo {author} {\bibfnamefont {C.}~\bibnamefont {Li}}, \bibinfo {author} {\bibfnamefont {J.}~\bibnamefont {Wang}}, \emph {et~al.},\ }\bibfield  {title} {\bibinfo {title} {Growth of single-crystal black phosphorus and its alloy films through sustained feedstock release},\ }\href {https://doi.org/10.1038/s41563-023-01516-1} {\bibfield  {journal} {\bibinfo  {journal} {Nat. Mater.}\ }\textbf {\bibinfo {volume} {22}},\ \bibinfo {pages} {717} (\bibinfo {year} {2023})}\BibitemShut {NoStop}%
\bibitem [{\citenamefont {Abramowitz}\ and\ \citenamefont {Stegun}(1972)}]{Abramowitz1972}%
  \BibitemOpen
  \bibfield  {author} {\bibinfo {author} {\bibfnamefont {M.}~\bibnamefont {Abramowitz}}\ and\ \bibinfo {author} {\bibfnamefont {I.~A.}\ \bibnamefont {Stegun}},\ }\href@noop {} {\emph {\bibinfo {title} {Handbook of mathematical functions}}},\ \bibinfo {edition} {10th}\ ed.\ (\bibinfo  {publisher} {National Bureau of Standards},\ \bibinfo {year} {1972})\ Chap.~\bibinfo {chapter} {20}\BibitemShut {NoStop}%
\end{thebibliography}%

\end{document}